\documentclass[10pt,aps,prc,twocolumn,superscriptaddress,floatfix]{revtex4-1}
\usepackage{amsmath,amssymb,mathrsfs}
\usepackage{bm}
\usepackage[dvipdfm]{graphicx}
\usepackage{ascmac}
\usepackage{geometry}
\def\bra<#1|{\left<#1\right|}
\def\ket|#1>{\left|#1\right>}
\def\braket<#1>{\left<#1\right>}
\def\bracket<#1|#2>{\left<#1\vphantom{#2}\right|\left.\kern-3pt\vphantom{#1}#2\right>}
\def\braccket<#1|#2|#3>{\left<#1\vphantom{#3}\right|#2\left|\vphantom{#1}#3\right>}
\def\lap{\bigtriangleup}
\def\sgn{\mathrm{sgn}}
\def\nonum\\{\nonumber\\}
\def\abs#1{\left|{#1}\right|}
\def\bs#1{\boldsymbol{#1}}

\def\bvecr{\boldsymbol{r}}

\def\RE#1#2+#3#4{\mbox{$^{#1}${#2}}+\mbox{$^{#3}${#4}}}
\def\setoftau{s(\{\tau_i\}:V_\mathrm{P}^nV_\mathrm{T}^{N-n})}

\begin{document}

\title{
Time-dependent Hartree-Fock calculations for multinucleon transfer processes
in $^{40,\,48}$Ca+$^{124}$Sn, $^{40}$Ca+$^{208}$Pb, and $^{58}$Ni+$^{208}$Pb reactions
}

\author{Kazuyuki Sekizawa}
\email[]{sekizawa@nucl.ph.tsukuba.ac.jp}
%\homepage[]{http://wwwnucl.ph.tsukuba.ac.jp/~sekizawa/}
\affiliation{Graduate School of Pure and Applied Sciences, University of Tsukuba, Tsukuba 305-8571, Japan}

\author{Kazuhiro Yabana}
\email[]{yabana@nucl.ph.tsukuba.ac.jp}
%\homepage[]{http://wwwnucl.ph.tsukuba.ac.jp/~yabana/}
\affiliation{Graduate School of Pure and Applied Sciences, University of Tsukuba, Tsukuba 305-8571, Japan}
\affiliation{Center for Computational Sciences, University of Tsukuba, Tsukuba 305-8577, Japan}

\date{June 27, 2013}

\begin{abstract}
Multinucleon transfer processes in heavy-ion reactions at energies 
slightly above the Coulomb barrier are investigated in a fully 
microscopic framework of the time-dependent Hartree-Fock (TDHF) 
theory. Transfer probabilities are calculated from the TDHF wave 
function after collision using the projection operator method 
which has recently been proposed by Simenel (C. Simenel, 
Phys. Rev. Lett. {\bf 105}, 192701 (2010)). We show results of 
the TDHF calculations for transfer cross sections of the reactions of
$^{40,\,48}$Ca+$^{124}$Sn at $E_\mathrm{lab}=$ 170, 174~MeV,  
$^{40}$Ca+$^{208}$Pb at $E_\mathrm{lab}=$ 235, 249~MeV, and
$^{58}$Ni+$^{208}$Pb at $E_\mathrm{lab}=$ 328.4~MeV, 
for which measurements are available. We find the transfer processes 
show different behaviors depending on the $N/Z$ ratios of the projectile 
and the target, and the product of the charge numbers, 
$Z_\mathrm{P}Z_\mathrm{T}$. When the projectile and the target 
have different $N/Z$ ratios, fast transfer processes of a few nucleons 
towards the charge equilibrium of the initial system occur in reactions 
at large impact parameters. As the impact parameter decreases, 
a neck formation is responsible for the transfer. A number of nucleons 
are transferred by the neck breaking when two nuclei dissociate, 
leading to transfers of protons and neutrons in the same direction.
Comparing cross sections by theory and measurements, 
we find the TDHF theory describes the transfer cross sections
of a few nucleons reasonably. As the number of transferred nucleons 
increases, the agreement becomes less accurate. The TDHF calculation 
overestimates transfer cross sections accompanying a large number 
of neutrons when more than one proton are transferred. 
Comparing our results with those by other theories, we find
the TDHF calculations give qualitatively similar results to those 
of direct reaction models such as GRAZING and Complex WKB. 
\end{abstract}


\maketitle

\section{INTRODUCTION}

In the last three decades, measurements of multinucleon transfer
processes have been achieved extensively in heavy-ion collisions
close to the Coulomb barrier \cite{MNT_EXP_1,MNT_EXP_2,
MNT_EXP_3,MNT_EXP_4,MNT_EXP_5,MNT_EXP_6,MNT_EXP_7,
MNT_EXP_8,MNT_EXP_9,Corradi(40Ca+124Sn),
Corradi(JPhysG1997),MNT_EXP_10,Corradi(48Ca+124Sn),
MNT_EXP_11,Corradi(64Ni+238U),MNT_EXP_12,
Corradi(58Ni+208Pb),Szilner(40Ca+208Pb)1,
Szilner(40Ca+208Pb)2,MNT_EXP_13,MNT_EXP_14,
MNT_EXP_15,vonOertzen(review),Corradi(review)}.
This process is an interesting quantum dynamics of many 
nucleons regarded as a nonequilibrium transport phenomenon 
which reflects both static properties and time-dependent dynamics 
of colliding nuclei. The static properties include nuclear shell 
structure and binding energies, nuclear shapes, properties of 
outermost single-particle orbitals, neutron-to-proton mass ratios, 
pairing properties, and so on. The dynamical effects include quantum 
tunneling between colliding nuclei, matching of $Q$-values and momenta, 
formation and breaking of the neck, quasi-fission dynamics, and so on.

Besides fundamental interests in its mechanisms, the multinucleon 
transfer reaction is also expected to be useful as a mean to produce 
unstable nuclei whose production is difficult by other methods. 
For example, a production of neutron-rich nuclei with atomic mass 
around $A \sim 200$ along the neutron magic number $N=126$ 
has been discussed \cite{Dasso(1994),Dasso(1995),Zagrebaev(2008),
Zagrebaev(2011)}. The knowledge on structural properties of these 
nuclei is crucially important to understand a detail scenario of heavy 
elements synthesis in the $r$-process. An experiment to produce such 
neutron-rich unstable nuclei has been planned in the reactions of Xe 
isotopes on $^{198}$Pt \cite{KISS}. A production of superheavy 
elements using multinucleon transfer reactions has also been 
discussed (see, for example, Refs.~\cite{Zagrebaev(2005),
Zagrebaev(2006),Zagrebaev(2007)1,Zagrebaev(2007)2,
Zagrebaev(2011),Simenel(review)}). 

To describe multinucleon transfer processes theoretically, 
models based on a direct reaction picture such as GRAZING 
\cite{GRAZING} and Complex WKB (CWKB) \cite{CWKB} have 
been extensively developed and applied \cite{MNT_EXP_9,
Corradi(40Ca+124Sn),Corradi(48Ca+124Sn),MNT_EXP_11,
Corradi(64Ni+238U),MNT_EXP_12,Corradi(58Ni+208Pb),
Szilner(40Ca+208Pb)1,Szilner(40Ca+208Pb)2,MNT_EXP_13}. 
In these models, multinucleon transfer processes are treated 
statistically, using single-nucleon transfer probabilities calculated 
by the first-order perturbation theory. A model based on 
Langevin-type equations of motion has also been developed 
\cite{Zagrebaev(2005),Zagrebaev(2007)1}. This model 
describes not only multinucleon transfer processes but also 
deep inelastic collisions, quasi-fission, fusion-fission, and fusion 
reactions in a unified way \cite{Zagrebaev(2005),
Zagrebaev(2006),Zagrebaev(2007)1,Zagrebaev(2007)2,
Zagrebaev(2008),Zagrebaev(2011)}. 

Although the above mentioned approaches have shown reasonable 
successes, these models are not fully microscopic but include some 
model assumptions. To get fundamental understanding of the 
dynamics and to present a reliable prediction for the cross sections,
it is highly desired to develop a fully microscopic description 
for the multinucleon transfer processes with minimum assumptions 
on the dynamics. To this end, we conduct microscopic calculations 
in the time-dependent Hartree-Fock (TDHF) theory.

The theory of the TDHF was first proposed by Dirac in 1930 
\cite{Dirac(TDHF)}. Applications of the TDHF theory to nuclear collision 
dynamics started in 1970s \cite{BKN(1976),Cusson(1976),
Koonin(1977),Davies(1978)1,Flocard(1978),Krieger(1978),Davies(1979),
Dasso(1979)}. Progresses in the early stage have been summarized in 
Ref.~\cite{Negele(review)}. Since then, continuous efforts have been 
devoted for improving the method and extending applications 
\cite{Umar(1986),Kim(1994),Kim(1997),Simenel(2003),
NakatsukasaYabana(2005),UO(liner-responce),Maruhn(GDR),UO(3D-FULL),
Maruhn(2006),UO(2006)1,Guo(2007),UO(2007),Guo(2008),Umar(2008),
Umar(2008)2,Washiyama(2008),Washiyama(2009),Golabek(2009),
Kedziora(2010)IQF,Iwata(2010),Projection,Iwata(2011),Keser(2012),
Iwata(2012),Simenel(QF2012)}. At present, three-dimensional 
calculations with full Skyrme functionals including time-odd 
components are routinely conducted. In most TDHF calculations, 
Skyrme-type interactions \cite{Skyrme(1956)} are used. Since 
parameters of Skyrme interactions are determined to reproduce 
nuclear properties for a wide mass region, there is no empirical 
parameter specific to the reaction.

The TDHF theory may describe both peripheral and central
collisions. In peripheral collisions, the mean-field of the 
collision partner works as a time-dependent perturbation for 
the orbitals. This picture of the transfer dynamics is similar 
to that in direct reaction models where single-particle transfer
probabilities are calculated either by the perturbation theory 
\cite{GRAZING,CWKB} or by solving numerically the 
time-dependent Schr{\"o}dinger equation \cite{
Bonaccorso(1987),Bonaccorso(1988),Bonaccorso(1991)}.
In collisions at smaller impact parameters, the TDHF theory
describes macroscopic dynamics such as fusion \cite{Krieger(1978),
Umar(1986),Kim(1997),UO(3D-FULL),UO(2007),Umar(2008)2,Keser(2012),
Simenel(review)}, quasi-fission \cite{Golabek(2009),Kedziora(2010)IQF,
Simenel(QF2012),Simenel(review)}, and deep inelastic collisions 
\cite{Cusson(1976),Koonin(1977),Davies(1978)1,Flocard(1978),
Davies(1979),Dasso(1979),Iwata(2010),Simenel(review)}.
Nucleons are exchanged between projectile and target nuclei
through the neck formation. This description of the multinucleon 
transfer processes is similar to the Langevin-type description 
\cite{Zagrebaev(2005),Zagrebaev(2007)1}.
In this way, the TDHF theory is expected to be capable of describing
quite different transfer mechanisms in a unified way.

In this paper, we will apply the TDHF theory to calculate 
transfer probabilities and cross sections for the reactions of 
\RE{40}{Ca}+{124}{Sn} at $E_\mathrm{lab}=$ 170~MeV, 
\RE{48}{Ca}+{124}{Sn} at $E_\mathrm{lab}=$ 174~MeV, 
\RE{40}{Ca}+{208}{Pb} at $E_\mathrm{lab}=$ 235, 249~MeV, and 
\RE{58}{Ni}+{208}{Pb} at $E_\mathrm{lab}=$ 328.4~MeV,
for which measurements are available
\cite{Corradi(40Ca+124Sn),Corradi(48Ca+124Sn),
Szilner(40Ca+208Pb)2,Corradi(58Ni+208Pb)}. 
To calculate transfer probabilities from the TDHF wave function 
after collision, we use the projection operator method which 
has recently been proposed by Simenel \cite{Projection}.

In addition to the fact that extensive measurements are available 
for these systems, analyses and comparisons of these systems 
are of much interest since reactions in these systems are expected 
to show qualitatively different features. 
While \RE{48}{Ca}+{124}{Sn} has almost the same 
neutron-to-proton ratio, $N/Z$, between the projectile 
and the target, other three systems have different $N/Z$ ratios. 
We expect transfer processes towards the charge equilibrium 
take place in collisions with large $N/Z$ asymmetry 
\cite{Freiesleben(1984),Iwata(2011),Iwata(2012)}. Moreover, 
it is well known that the basic feature of the low-energy heavy-ion 
collisions depends much on the product of the charge numbers of 
the projectile and the target nuclei, $Z_\mathrm{P}Z_\mathrm{T}$.
Fusion reactions beyond the critical value, 
$Z_\mathrm{P}Z_\mathrm{T} \sim 1600$, are known to accompany 
an extra-push energy \cite{Extra-push(1984)EXP,Simenel(review)}. 
Analyses of the fusion-hindrance phenomena in the TDHF theory have 
been reported in Ref.~\cite{Simenel(review)}, showing that the extra-push 
energy in the TDHF calculation is in good agreement with that of the 
Swiatecki's extra-push model \cite{extra-push}. The four systems to 
be analyzed have different $Z_\mathrm{P}Z_\mathrm{T}$ values, 
1000 for \RE{40,\,48}{Ca}+{124}{Sn}, 1640 for 
\RE{40}{Ca}+{208}{Pb}, and 2296 for \RE{58}{Ni}+{208}{Pb}.

It has been considered that the success of the TDHF theory is limited to 
observables expressed as expectation values of one-body operators. 
Indeed, the particle number fluctuation in deep inelastic 
collisions has been found to be substantially underestimated in 
the TDHF calculations \cite{Koonin(1977),Davies(1978)1,Dasso(1979)}. 
Since transfer probabilities in the TDHF calculation may not be given as 
expectation values of any one-body operators, it is not at all obvious 
whether the TDHF calculation provides a reasonable description for 
multinucleon transfer processes. One of the main purposes 
of the present paper is to clarify usefulness and limitation of the TDHF 
calculation for the multinucleon transfer processes. We note that 
Simenel has recently presented a calculation using the Barian-V\'en\'eroni 
prescription \cite{BV(1981)} and concluded that the particle number 
fluctuation may not be affected much by the correlation effects beyond 
the TDHF theory for reactions which are not so much violent as deep 
inelastic collisions \cite{Simenel(BV2011)}.

The construction of this paper is as follows. In Section 
\ref{Sec:Formulation}, we describe a formalism to calculate transfer 
probabilities from the TDHF wave function after collision. We also 
describe our computational method. In Section \ref{Sec:Results}, 
we present results of our TDHF calculations for four systems and 
compare them with measurements. In Section \ref{Sec:Comparison}, 
we compare our results with those by other theories. In Section 
\ref{Sec:Summary_and_Conclusions}, a summary and a future 
prospect will be presented.

\section{FORMULATION}{\label{Sec:Formulation}}

\subsection{Definition of transfer probabilities}

We consider a collision of two nuclei described by the TDHF theory. 
The projectile is composed  of $N_\mathrm{P}$ nucleons and 
the target is composed of $N_\mathrm{T}$ nucleons. The total 
number of nucleons is $N=N_\mathrm{P}+N_\mathrm{T}$. 
In the TDHF calculation, a time evolution of single-particle orbitals, 
$\phi_i(\bvecr,\sigma,t)$ ($i=1,\cdots,N$), is calculated where 
$\bvecr$ and $\sigma$ denote the spatial and the spin coordinates, 
respectively. The total wave function is given by the Slater determinant 
composed of the orbitals:
\begin{equation}
\Phi(x_1,\cdots,x_N,t) =
\frac{1}{\sqrt{N!}}\det\bigl\{ \phi_i(x_j,t) \bigr\},
\label{Phi_Slater}
\end{equation}
where $x$ is a set of the spatial and the spin coordinates, 
$x\equiv$ $(\bvecr,\sigma)$. For the moment, we will develop a 
formalism for a many-body system composed of identical fermions. 
An extension to the actual nuclei composed of two kinds of 
fermions, protons and neutrons, is simple and obvious.

Before the collision, two nuclei are separated spatially.
We divide the whole space into two, the projectile region, 
$V_\mathrm{P}^i$, and the target region, $V_\mathrm{T}^i$. 
After the collision, we assume that there appear two nuclei, 
a projectile-like fragment (PLF) and a target-like fragment (TLF). 
We ignore channels in which nuclei are separated into more than 
two fragments after the collision. We again introduce a division 
of the whole space into two, the projectile region, 
$V_\mathrm{P}^f$, which includes the PLF, and the target region, 
$V_\mathrm{T}^f$, which includes the TLF.

We define the number operator of each spatial region as
\begin{equation}
\hat{N}_\tau = 
\int_\tau d\bvecr \;\sum_{i=1}^N \delta(\bvecr - \bvecr_i) = 
\sum_{i=1}^N \Theta_\tau(\bvecr_i),
\label{hat_N_tau}
\end{equation}
where $\tau$ specifies the spatial region either 
$V_\mathrm{P}^{i(f)}$ or $V_\mathrm{T}^{i(f)}$.
We introduce the space division function, $\Theta_\tau(\bvecr)$, 
defined as
\begin{equation}
\Theta_\tau(\bvecr) = \biggl\{
\begin{array}{ccc}
1 & \mathrm{for} & \bvecr \in \tau \\
0 & \mathrm{for} & \bvecr \notin \tau
\end{array}.
\label{SDF}
\end{equation}
The sum of the two operators, 
$\hat{N}_{V_\mathrm{P}^{i(f)}}$ and 
$\hat{N}_{V_\mathrm{T}^{i(f)}}$, is the number operator 
of the whole space, $\hat{N}=$ $\hat{N}_{V_\mathrm{P}^{i}}
+\hat{N}_{V_\mathrm{T}^{i}}$ $=\hat{N}_{V_\mathrm{P}^{f}}
+\hat{N}_{V_\mathrm{T}^{f}}$. In ordinary TDHF calculations, 
an initial wave function is the direct product of the ground state wave 
functions of two nuclei boosted with the relative velocity. The 
single-particle orbitals, $\phi_i(x,t)$, are localized in one of the 
spatial regions, $V_\mathrm{P}^i$ or $V_\mathrm{T}^i$, 
at the initial stage of the calculation. Therefore, the initial wave 
function is the eigenstate of both operators, 
$\hat{N}_{V_\mathrm{P}^i}$ and 
$\hat{N}_{V_\mathrm{T}^i}$, with eigenvalues, 
$N_\mathrm{P}$ and $N_\mathrm{T}$, respectively. 
At the final stage of the calculation after the collision, 
each single-particle orbital extends spatially to both spatial 
regions of $V_\mathrm{P}^f$ and $V_\mathrm{T}^f$. 
Due to this fact, the Slater determinant at the final stage is 
not an eigenstate of the number operators, 
$\hat{N}_{V_\mathrm{P}^f}$ and 
$\hat{N}_{V_\mathrm{T}^f}$, but a superposition of 
states with different particle number distributions.

The probability that $n$ nucleons are in the spatial region 
$V_\mathrm{P}^f$ and $N-n$ nucleons are in the 
spatial region $V_\mathrm{T}^f$ is defined as follows.
We start with the normalization relation of the final wave function 
after the collision,
\begin{equation}
\int\hspace{-0.5mm} dx_1 \cdots \int\hspace{-0.8mm} dx_N 
\abs{\Psi(x_1,\cdots,x_N)}^2 = 1,
\label{normalization1}
\end{equation}
where $\int dx \equiv \sum_\sigma \int d\bvecr$.
Here and hereafter, we denote the many-body wave function at the 
final stage of the calculation as $\Psi(x_1,\cdots,$ $x_N)=$ 
$\det\{ \psi_i(x_j)\}/\sqrt{N!}$, and omit the time 
index. We also omit the suffix $f$ from $V_\mathrm{P}^f$ 
and $V_\mathrm{T}^f$. The normalization relation, 
Eq.~(\ref{normalization1}), includes $N$-fold integral over the whole 
spatial region. We divide each spatial integral into two integrals over 
the subspaces, $V_\mathrm{P}$ and $V_\mathrm{T}$. 
We then classify the $2^N$ terms, generated by the divisions of the 
spatial regions, according to the number of $V_\mathrm{P}$ and 
the number of $V_\mathrm{T}$ included in the integral:
\begin{eqnarray}
1 &=& 
\int_{V_\mathrm{P}+V_\mathrm{T}}\hspace{-6mm}dx_1 \cdots 
\int_{V_\mathrm{P}+V_\mathrm{T}}\hspace{-6mm}dx_N
\abs{\Psi(x_1,\cdots,x_N)}^2 \nonum\\
&=& \hspace{-0.2mm}
\sum_{n=0}^N \sum_{\setoftau}\hspace{-0.2mm}
\int_{\tau_1}\hspace{-2.5mm}dx_1 \hspace{-0.5mm}\cdots\hspace{-1.5mm} 
\int_{\tau_N}\hspace{-3.5mm}dx_N \hspace{-0.6mm}
\abs{\Psi(x_1,\cdots,x_N)}^2 \hspace{-1.5mm},\nonum\\[-2mm]
\label{normalization2}
\end{eqnarray}
where each subscript $\tau_i$ ($i=1,\cdots,N$) represents either 
$V_\mathrm{P}$ or $V_\mathrm{T}$. The notation $\setoftau$ 
means that the sum should be taken for all possible combinations 
of $\tau_i$ on condition that, in the sequence of 
$\tau_1,\cdots,\tau_N$, $V_\mathrm{P}$ appears $n$ times 
and $V_\mathrm{T}$ appears $N-n$ times. The number of the 
combinations equals to $_NC_n$. From this expression, we find 
the probability that $n$ nucleons are in the $V_\mathrm{P}$ and 
$N-n$ nucleons are in the $V_\mathrm{T}$ is given by
\begin{equation}
P_n = \hspace{-6mm}
\sum_{\setoftau} \hspace{-0.5mm}
\int_{\tau_1}\hspace{-1.5mm}dx_1 \cdots\hspace{-0.5mm} 
\int_{\tau_N}\hspace{-2.5mm}dx_N \hspace{-0.5mm}
\abs{\Psi(x_1,\cdots,x_N)}^2.
\label{Definition_of_P_n}
\end{equation}
Equation (\ref{normalization2}) ensures the relation, 
$\sum_{n=0}^N P_n = 1$. From the probability $P_n$, we may 
obtain nucleon transfer probabilities. For example, the probability 
of $n$-particle transfer from the projectile to the target is given 
by $P_{N_\mathrm{P}-n}$.

\subsection{Number projection operator}

Above expression of the probability $P_n$ can be represented as an 
expectation value of the number projection operator 
$\hat{P}_n$, {\it i.e.} $P_n=\braccket<\Psi|\hat{P}_n|\Psi>$.
This operator extracts a component of the wave function with particle 
number $n$ in the $V_\mathrm{P}$ and $N-n$ in the $V_\mathrm{T}$ 
from the final wave function $\Psi(x_1,\cdots,x_N)$.
From Eq.~(\ref{Definition_of_P_n}), we obtain the following expression 
for the number projection operator, 
\begin{equation}
\hat{P}_n = \hspace{-4mm}
\sum_{\setoftau}\hspace{-4mm}
\Theta_{\tau_1}(\bvecr_1)\cdots\Theta_{\tau_N}(\bvecr_N).
\label{projection_op_Theta}
\end{equation}
The projected wave function, $\hat{P}_n \Psi$, is the eigenstate 
of the number operators, $\hat{N}_{V_\mathrm{P}}$ and 
$\hat{N}_{V_\mathrm{T}}$, with eigenvalues, $n$ and $N-n$, 
respectively. From Eq.~(\ref{normalization2}), there follows
\begin{equation}
\sum_{n=0}^N \hat{P}_n = \prod_{i=1}^N \Bigl( 
\Theta_{V_\mathrm{P}}(\bvecr_i) + 
\Theta_{V_\mathrm{T}}(\bvecr_i) \Bigr) = 1.
\end{equation}

Recently, Simenel has provided an alternative expression for the
number projection operator \cite{Projection} which is given by
\begin{equation}
\hat{P}_n = 
\frac{1}{2\pi}\int_{0}^{2\pi}\hspace{-2mm}d\theta\, 
e^{i(n-\hat{N}_{V_\mathrm{P}})\theta}.
\label{projection_op_phase}
\end{equation}
We can easily show that this expression, 
Eq.~(\ref{projection_op_phase}), is equivalent to 
Eq.~(\ref{projection_op_Theta}) as follows:
\begin{eqnarray}
\hat{P}_n &=& 
\frac{1}{2\pi}\int_{0}^{2\pi}\hspace{-2mm}d\theta\, 
e^{i(n-\hat{N}_{V_\mathrm{P}})\theta} \nonum\\
&=& \frac{1}{2\pi}\int_{0}^{2\pi}\hspace{-2mm}d\theta\, 
e^{in\theta}\,\prod_{i=1}^{N}
\Bigl( \Theta_{V_\mathrm{T}}(\bvecr_i) 
+ e^{-i\theta}\Theta_{V_\mathrm{P}}(\bvecr_i) \Bigr) \nonum\\
&=& \sum_{n'=0}^{N} 
\frac{1}{2\pi}\int_{0}^{2\pi}\hspace{-2mm} 
e^{i(n-n')\theta}d\theta \hspace{-8mm}\sum_
{s(\{\tau_i\}:V_\mathrm{P}^{n'}V_\mathrm{T}^{N-n'})} 
\hspace{-6mm}
\Theta_{\tau_1}(\bvecr_1)\cdots
\Theta_{\tau_N}(\bvecr_N) \nonum\\[1mm]
&=& \hspace{-4mm}
\sum_{\setoftau}\hspace{-4mm}
\Theta_{\tau_1}(\bvecr_1)\cdots
\Theta_{\tau_N}(\bvecr_N). \nonumber
\label{equivalence_of_projection_op}
\end{eqnarray}

\subsection{Computation of transfer probabilities}

Two expressions for the number projection operator $\hat P_n$,
Eq.~(\ref{projection_op_Theta}) and Eq.~(\ref{projection_op_phase}),
have been utilized to calculate transfer probabilities in the TDHF theory. 
When we use Eq.~(\ref{projection_op_Theta}), the probability $P_n$ 
is expressed in terms of the single-particle orbitals as
\begin{eqnarray}
P_n  
&=& \hspace{-0.7mm}
\int\hspace{-1.2mm}dx_1 \cdots\hspace{-1.3mm}
\int\hspace{-1.1mm}dx_N\,
\psi_1^*(x_1)\cdots\psi_N^*(x_N) \hat{P}_n
\det\bigl\{ \psi_i(x_j) \bigr\} \nonum\\
&=& \hspace{-4mm}
\sum_{\setoftau}\hspace{-0.5mm}\sum_{\xi}\sgn(\xi)
\bracket<\psi_1|\psi_{\xi_1}>_{\tau_1} \hspace{-1mm}\cdots 
\bracket<\psi_N|\psi_{\xi_N}>_{\tau_N} \nonum\\
&=& \hspace{-4mm}
\sum_{\setoftau}\hspace{-5mm}
\det\bigl\{ \bracket<\psi_i|\psi_j>_{\tau_i} \bigr\},
\label{P_n_Slater}
\end{eqnarray}
where the summation over $\xi$ is taken for all possible 
permutations of the index $\xi_i$ ($i=1,\cdots,N$), and 
$\sgn(\xi)$ is a sign depending on the number of permutations. 
$\bracket<\psi_i|\psi_j>_\tau \equiv$ 
$\int_\tau dx\, \psi_i^*(x)\psi_j(x)$ 
denotes an overlap integral in the spatial region $\tau$.

When we use Eq.~(\ref{projection_op_phase}), we obtain
\begin{eqnarray}
P_n &=& 
\frac{1}{2\pi}\int_{0}^{2\pi}\hspace{-2mm}d\theta\, 
e^{in\theta}\,\braccket<\Psi|\prod_{i=1}^{N}
e^{-i\Theta_{V_\mathrm{P}}(\bvecr_i)\theta}|\Psi>\nonum\\
&=& \frac{1}{2\pi}\int_{0}^{2\pi}\hspace{-2mm}d\theta\, 
e^{in\theta}\,\det\bigl\{ \bracket<\psi_i|\psi_j>_{V_\mathrm{T}}
+e^{-i\theta}\bracket<\psi_i|\psi_j>_{V_\mathrm{P}} \bigr\}. 
\nonum\\[2mm]\nonum\\[-5mm]
\label{P_n_projection}
\end{eqnarray}

Two expressions, Eq.~(\ref{P_n_Slater}) and Eq.~(\ref{P_n_projection}), 
should give equivalent results. We indeed confirmed that both expressions 
give the same results for light systems. However, the computational cost is 
rather different between two methods. Let us first consider the 
computational cost of Eq.~(\ref{P_n_Slater}),
\begin{equation*}
P_n = \hspace{-4mm}\sum_{\setoftau}\hspace{-5mm}
\det\bigl\{ \bracket<\psi_i|\psi_j>_{\tau_i} \bigr\}.
\end{equation*}
In this expression, it is necessary to calculate the determinants
of dimension $N$ many times. For example, to calculate the
probabilities of all possible processes, $P_0$~to~$P_N$, we need
to calculate determinants of dimension $N$ for $2^N$ times.
Even for the calculation of the probability without any particle 
transfer, we need to calculate the determinants as many as 
$_NC_{N_\mathrm{P}}$. The calculation in this way soon becomes
impossible as $N$ increases and is useful only for light systems. 
This method has been used in the
\RE{40}{Ca}+{40}{Ca} collision in Ref.~\cite{Koonin(1977)}. 
It has also been used in the electron transfer processes in atomic 
collisions \cite{Method1983,Nagano(2000)}.

When we use the expression of Eq.~(\ref{P_n_projection}),
\begin{equation*}
P_n = 
\frac{1}{2\pi}\int_{0}^{2\pi}\hspace{-2mm}d\theta\, 
e^{in\theta}\,\det\bigl\{ \bracket<\psi_i|\psi_j>_{V_\mathrm{T}}
+e^{-i\theta}\bracket<\psi_i|\psi_j>_{V_\mathrm{P}} \bigr\},
\end{equation*}
the computational cost can be significantly small. In this expression, 
we achieve integral over $\theta$ employing the trapezoidal rule 
discretizing the interval [0,~2$\pi$] into $M$ equal grids. To calculate all the 
probabilities, $P_0$~to~$P_N$, we need to calculate the determinants 
of dimension $N$ for $M$ times. We find $M=200$ is sufficient for 
systems presented in this paper. In our calculations shown below, 
we employ Eq.~(\ref{P_n_projection}).

\subsection{Transfer cross sections}{\label{Subsec:x-sections}}

We next derive the formula for cross sections of transfer 
reactions. We assume that both projectile and target nuclei 
are spherical, so that the reaction is specified by the incident 
energy $E$ and the impact parameter $b$.

Up to this point, we derived expressions of transfer probabilities 
for a system composed of identical fermions. Since the TDHF 
wave function is a direct product of Slater determinants for 
protons and neutrons, the reaction probability is also given 
by the product of the probabilities for protons and neutrons. 
Let us denote the probability that $Z$ protons are included 
in the $V_\mathrm{P}$ as $P_Z^{(p)}(b)$ and $N$ 
neutrons are included in the $V_\mathrm{P}$ as 
$P_N^{(n)}(b)$. Then, the probability that $Z$ protons 
and $N$ neutrons are included in the $V_\mathrm{P}$ 
is given by
\begin{equation}
P_{Z,N}(b) = P_Z^{(p)}(b)\,P_N^{(n)}(b).
\end{equation}
We calculate the transfer cross section for the channel where 
the PLF is composed of $Z,N$ nucleons by integrating the 
probability $P_{Z,N}(b)$ over the impact parameter,
\begin{equation}
\sigma_\mathrm{tr}(Z,N) = 
2\pi\int_{b_\mathrm{min}}^{\infty}\hspace{-2.5mm}
b\; P_{Z,N}(b)\,db.
\label{x-section}
\end{equation}
The minimum of the integration over the impact parameter
is the border dividing fusion and binary reactions. 
In practice, we first examine the maximum impact parameter 
in which fusion reactions take place for a given incident energy.
We will call it the fusion critical impact parameter and
denote it as $b_\mathrm{f}$. We then repeat reaction calculations 
at various impact parameters for the region, $b>b_\mathrm{f}$, 
and calculate the cross section by numerical quadrature according 
to Eq.~(\ref{x-section}).

\subsection{Numerical methods}{\label{Subsec:Numerical_methods}}

We have developed our own computational code of the TDHF theory 
for heavy-ion collisions extending the code developed for the real-time
linear response calculations \cite{NakatsukasaYabana(2005)}. We 
employ a uniform spatial grid in the three-dimensional Cartesian 
coordinate to represent single-particle orbitals without any symmetry 
restrictions. The grid spacing is taken to be 0.8~fm. We take a box size 
of $60 \times 60 \times 26$ grid points (48~fm $\times$ 48~fm 
$\times$ 20.8~fm) for collision calculations, where the reaction plane is 
taken to be the $xy$-plane. The initial wave functions of projectile and 
target nuclei are prepared in a box with $26 \times 26 \times 26$ grid 
points. We use 11-points finite-difference formula for the first and second 
derivatives. To calculate the time evolution of single-particle orbitals, we use 
the Taylor expansion method of 4th order. The first-order predictor-corrector 
step is adopted in the time evolution. The time step is set to 
$\Delta t=0.2$~fm/c. To calculate the Coulomb potential, we employ 
the Hockney's method \cite{Hockney} in which the Fourier transformation 
is achieved in the grid of two times larger box than that utilized to express 
single-particle orbitals.

We have tested the accuracy of the code by 
comparing our results with those by other codes. We have confirmed 
that the fusion critical impact parameters of the reactions of 
\RE{16}{O}+{16}{O} and \RE{16}{O}+{28}{O} reported in 
Ref.~\cite{UO(3D-FULL)} are reproduced within 0.1~fm accuracy by 
our code. We have also calculated the fluctuation of exchanged 
nucleons for \RE{40}{Ca}+{40}{Ca} head-on collisions and 
confirmed that results reported in Ref.~\cite{Washiyama(2009)SMF} 
are reproduced accurately.

\section{RESULTS}{\label{Sec:Results}}

In this section, we will show calculated results for the reactions of 
\RE{40}{Ca}+{124}{Sn} at the incident energy of 170~MeV, 
\RE{48}{Ca}+{124}{Sn} at 174~MeV,
\RE{40}{Ca}+{208}{Pb} at 235, 249~MeV, and 
\RE{58}{Ni}+{208}{Pb} at 328.4~MeV. 

As for the energy density functional and potential, we use the Skyrme 
functional including all time-odd terms \cite{Full-Skyrme} except for the 
second derivative of the spin densities, $\lap{\bs{s}^{(n,\,p)}}$. 
We encounter numerical instability in the time evolution calculation 
if we include the term in the potential. All of the results reported here 
are calculated using the Skyrme SLy5 parameter set \cite{Chabanat}. 
This interaction has been utilized in the fully three-dimensional TDHF 
calculations for heavy-ion collisions \cite{UO(3D-FULL),Umar(2008)2,Keser(2012)}.

In the ground state calculations, we find the ground states of $^{40}$Ca, 
$^{48}$Ca and $^{208}$Pb are spherical. The ground state of $^{124}$Sn 
is oblately deformed with $\beta \sim$ 0.11. The ground state of 
$^{58}$Ni is prolately deformed with $\beta \sim$ 0.11.

We take the incident direction parallel to the $x$-axis and the impact
parameter vector parallel to the $y$-axis. The reaction is specified by 
the incident energy and the impact parameter.  As an initial condition, 
the two colliding nuclei are placed with the distance 16-18~fm in the 
$x$-direction. Before starting the TDHF calculation, we assume the
centers of the two colliding nuclei follow the Rutherford trajectory.
For the deformed nuclei, we placed the nucleus with the symmetry 
axis being set parallel to the $z$-axis.

We stop time evolution calculations when two nuclei are separated 
by 20-26~fm, if binary fragments are produced. If the colliding nuclei 
fuse and do not separate, we continue time evolution calculations 
more than 3000~fm/c after two nuclei touch. We have not found any 
reactions in which more than two fragments are produced after collision.

For each collision system, we first find the fusion critical 
impact parameter $b_\mathrm{f}$. We find them by repeating 
calculations changing the impact parameter by 0.01~fm step. We 
then calculate reactions for various impact parameters outside the 
critical value. At an impact parameter region smaller than 7~fm, we 
calculate reactions of impact parameters with 0.25~fm step. 
At an impact parameter region larger than 7~fm, we calculate 
reactions of $b=$ 7.5, 8, 9, and 10~fm. Close to the fusion critical 
impact parameter, we calculate reactions in 0.05~fm and 0.01~fm 
impact parameter steps. All these calculations are used to evaluate 
the transfer cross sections. In calculating transfer cross sections 
according to Eq.~(\ref{x-section}), the upper limit of the 
integral over $b$ is set to 10~fm.

\subsection{\RE{40,\,48}{Ca}+{124}{Sn} reactions}

In this subsection, we present results for the reactions 
of \RE{40}{Ca}+{124}{Sn} at $E_\mathrm{lab}=$ 170~MeV 
($E_\mathrm{c.m.}\simeq$ 128.5~MeV) and 
\RE{48}{Ca}+{124}{Sn} at $E_\mathrm{lab}=$ 174~MeV
($E_\mathrm{c.m.}\simeq$ 125.4~MeV), for which multinucleon 
transfer cross sections have been measured experimentally 
\cite{Corradi(40Ca+124Sn), Corradi(48Ca+124Sn)}.
The neutron-to-proton ratio, $N/Z$, is different between the 
projectile and the target for \RE{40}{Ca}+{124}{Sn}, while it is 
almost the same for \RE{48}{Ca}+{124}{Sn}. Therefore, 
we may expect different features in the transfer process. 
As we mentioned in the introduction, the product of charge
numbers of the projectile and the target is important for the 
fusion dynamics. The present systems have 
$Z_\mathrm{P}Z_\mathrm{T}=$ 1000 $<$ 1600, so that no 
fusion-hindrance is expected to occur.

To estimate the Coulomb barrier height, we calculate the 
nucleus-nucleus potential using the frozen-density approximation 
neglecting the Pauli blocking effect \cite{Brueckner(1968),
Denisov(2002),Washiyama(2008)}. The potential is given by 
$V(R)=$ $E[\rho_\mathrm{{}_P}+\rho_\mathrm{{}_T}](R)
-E_\mathrm{g.s.}[\rho_\mathrm{{}_P}]-E_\mathrm{g.s.}
[\rho_\mathrm{{}_T}]$, where $R$ is the distance between the 
centers-of-masses of the two nuclei, $\rho_\mathrm{{}_P}$ 
($\rho_\mathrm{{}_T}$) denotes nuclear density of the projectile 
(target) in their ground state. $E[\rho_\mathrm{{}_P}+
\rho_\mathrm{{}_T}](R)$ denotes the total energy when two nuclei 
are separated by the relative distance $R$. 
$E_\mathrm{g.s.}[\rho_\mathrm{{}_P}]$ and 
$E_\mathrm{g.s.}[\rho_\mathrm{{}_T}]$ denote the ground 
state energy of each nucleus. In the calculation, the Coulomb 
barrier height is estimated as $V_\mathrm{B} \approx$ 116.3~MeV 
for \RE{40}{Ca}+{124}{Sn} and $V_\mathrm{B} \approx$ 
115.1~MeV for \RE{48}{Ca}+{124}{Sn}, respectively. Since the 
initial relative energies are higher than the Coulomb 
barrier heights, we find the fusion critical impact parameter, 
$b_\mathrm{f}$ = 3.95~fm for \RE{40}{Ca}+{124}{Sn} and 
$b_\mathrm{f}$ = 3.93~fm for \RE{48}{Ca}+{124}{Sn}, respectively.

\subsubsection{Overview of the reactions}

%&&&&&&&&&&&&&&&&&&&&&&&&&&&&&&&&&&&&&&&&&&&&&&&&&&
\begin{figure}[t]
   \begin{center}
   \includegraphics[width=6.5cm]{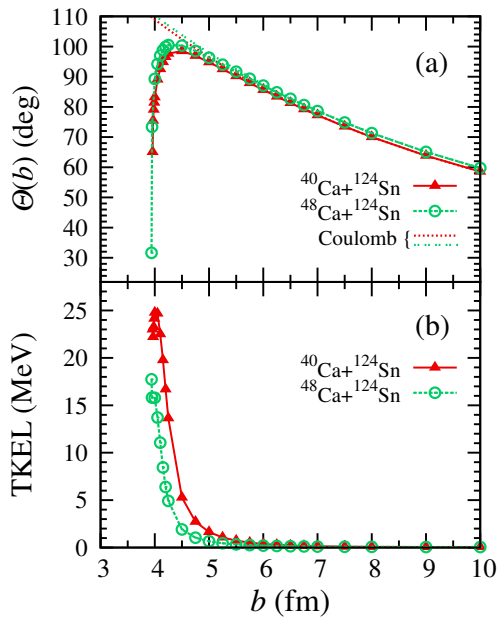}
   \end{center}\vspace{-3mm}
   \caption{(Color online)
   Deflection function (a) and total kinetic energy loss (b) as functions of 
   impact parameter $b$ for the reactions of \RE{40}{Ca}+{124}{Sn} at 
   $E_\mathrm{lab}=$ 170~MeV and \RE{48}{Ca}+{124}{Sn} at
   $E_\mathrm{lab}=$ 174~MeV. Results for the \RE{40}{Ca}+{124}{Sn} 
   reactions are denoted by red filled triangles connected with solid lines, while 
   results for the \RE{48}{Ca}+{124}{Sn} reactions are denoted by 
   green open circles connected with dashed lines. In (a), we also show 
   deflection functions for the pure Coulomb trajectories by a red dotted 
   line for the \RE{40}{Ca}+{124}{Sn} reactions and by a green two-dot chain 
   line for the \RE{48}{Ca}+{124}{Sn} reactions.
   }
   \label{FIG:Theta+TKEL(40,48Ca+124Sn)}
\end{figure}
%&&&&&&&&&&&&&&&&&&&&&&&&&&&&&&&&&&&&&&&&&&&&&&&&&&

Before showing detailed analyses of transfer reactions, we first
present an overview of the reaction dynamics.
In Fig.~\ref{FIG:Theta+TKEL(40,48Ca+124Sn)}, we show the 
deflection function, $\Theta(b)$, in (a) and the total kinetic energy 
loss (TKEL) in (b), as functions of impact parameter $b$. Results 
for the \RE{40}{Ca}+{124}{Sn} reactions are denoted by red 
filled triangles connected with solid lines, while results for the 
\RE{48}{Ca}+{124}{Sn} reactions are denoted by green open 
circles connected with dotted lines. 
In Fig.~\ref{FIG:Theta+TKEL(40,48Ca+124Sn)} (a), we also show 
deflection functions of the pure Coulomb trajectories by a red 
dotted line for \RE{40}{Ca}+{124}{Sn} and by a green two-dot 
chain line for \RE{48}{Ca}+{124}{Sn}. 

In practice, the deflection function and the TKEL are calculated
in the following way. We denote the center-of-mass coordinate 
of the PLF (TLF) and the relative coordinate as 
$\bs{R}_\mathrm{PLF\,(TLF)}(t)$ and
$\bs{R}(t) = \bs{R}_\mathrm{PLF}(t) - \bs{R}_\mathrm{TLF}(t)$,
respectively. We also denote  the mass, charge number, and the reduced 
mass at the final stage of the calculation as $M_\mathrm{PLF\,(TLF)}$, 
$Z_\mathrm{PLF\,(TLF)}$, and $\mu_f = M_\mathrm{PLF}M_\mathrm{TLF}/
(M_\mathrm{PLF}+M_\mathrm{TLF})$. The relative velocity at the 
final stage of the calculation, $t=t_f$, is calculated by 
$\dot{\bs{R}}(t_f) = (\bs{R}(t_f+\Delta t)-\bs{R}(t_f-\Delta t))/2\Delta t$.
We evaluate the TKEL by $\mathrm{TKEL} = E_\mathrm{c.m.} 
- \frac{1}{2} \mu_f \dot{\bs{R}}(t_f)^2 
-Z_\mathrm{PLF}Z_\mathrm{TLF}\,e^2/\abs{\bs{R}(t_f)}$, where 
$E_\mathrm{c.m.}$ is the initial incident energy in the center-of-mass frame. 
The angle between the vector $\bs{R}(t_f)$ and the $x$-axis, or the angle 
between the vector $\dot{\bs{R}}(t_f)$ and the $x$-axis, provides 
approximate value of the deflection angle. We estimate the correction for it
assuming that both the PLF and the TLF follow the Rutherford trajectory 
specified by the coordinates and the velocities at the final time, $t_f$.

%&&&&&&&&&&&&&&&&&&&&&&&&&&&&&&&&&&&&&&&&&&&&&&&&&&
\begin{figure}[t]
   \begin{center}
   \includegraphics[width=8cm]{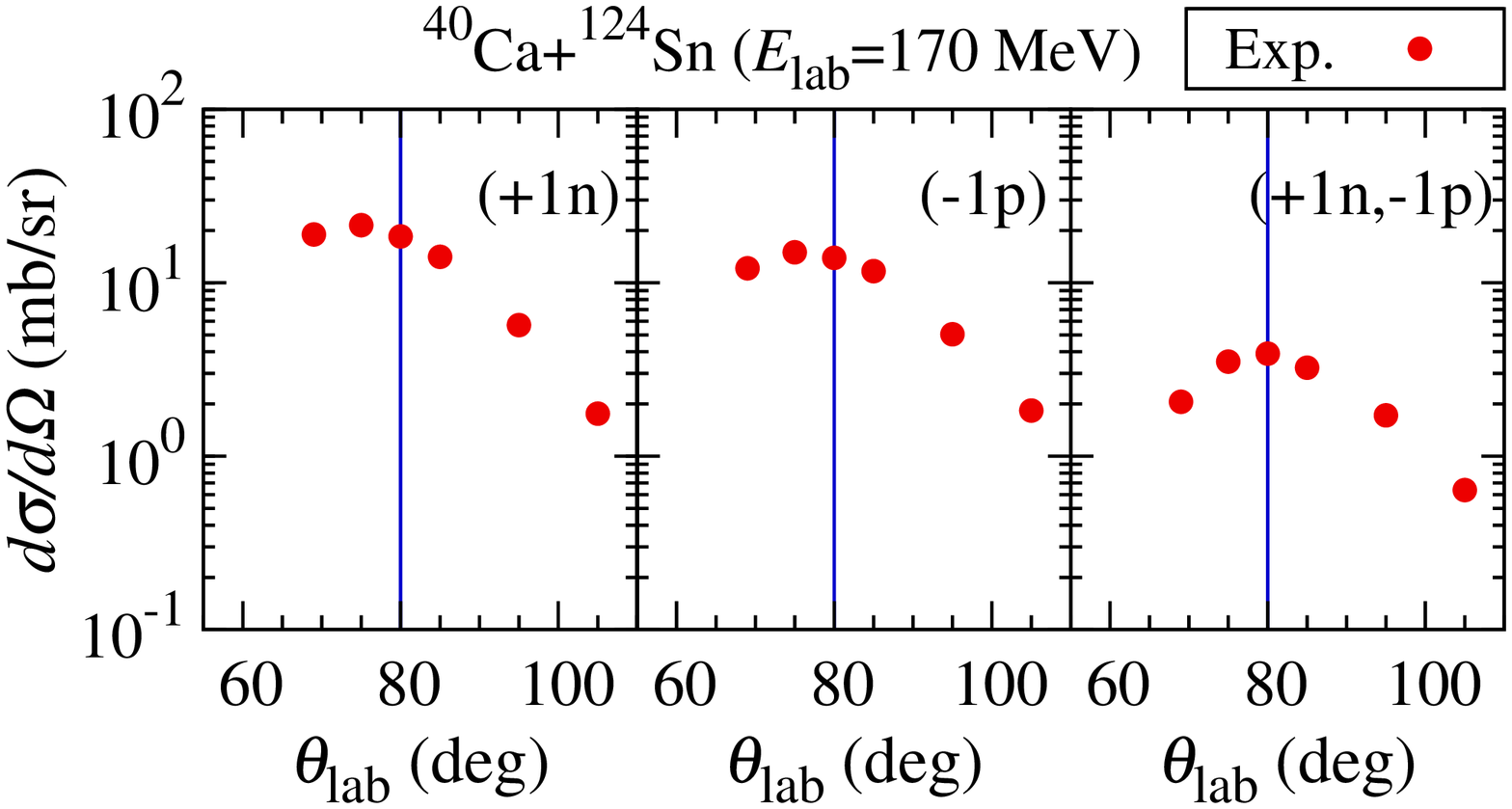}
   \end{center}\vspace{-3mm}
   \caption{(Color online)
   Differential cross sections of representative transfer 
   channels as functions of scattering angle in the laboratory 
   frame for the \RE{40}{Ca}+{124}{Sn} reaction at 
   $E_\mathrm{lab}=$ 170~MeV. 
   The Coulomb rainbow angle obtained from the TDHF trajectories 
   is denoted by blue solid vertical lines, and is compared with
   measured differential cross sections, red filled circles, 
   which have been reported in Ref.~\cite{Corradi(40Ca+124Sn)}. 
   }
   \label{FIG:DCS(40Ca+124Sn)_exp}
\end{figure}
%&&&&&&&&&&&&&&&&&&&&&&&&&&&&&&&&&&&&&&&&&&&&&&&&&&

The TKEL increases rapidly as the impact parameter decreases in the 
region $b < 4.5$~fm, where the deflection function, $\Theta(b)$, 
decreases appreciably by the nuclear attractive interaction. 
The deflection function shows a maximum at $b \sim$ 4.25~fm and 
decreases inside this impact parameter. The maximum 
	deflection angle corresponds to the Coulomb rainbow angle, 
$\theta_\mathrm{r}$. It is given by $99^\circ$ for 
\RE{40}{Ca}+{124}{Sn} and $100^\circ$ for 
\RE{48}{Ca}+{124}{Sn}. In Fig.~\ref{FIG:DCS(40Ca+124Sn)_exp}, 
we compare the Coulomb rainbow angle for the \RE{40}{Ca}+{124}{Sn} 
reaction with measured differential cross sections reported in 
Ref.~\cite{Corradi(40Ca+124Sn)}. Red filled circles denote measured 
cross sections and blue solid vertical lines denote the Coulomb rainbow 
angle in the laboratory frame. As seen from the figure, the peak positions 
of the measured cross sections roughly coincide with the Coulomb 
rainbow angle by the TDHF calculation.

%&&&&&&&&&&&&&&&&&&&&&&&&&&&&&&&&&&&&&&&&&&&&&&&&&&
\begin{figure}[b]
   \begin{center}
   \includegraphics[width=8.6cm]{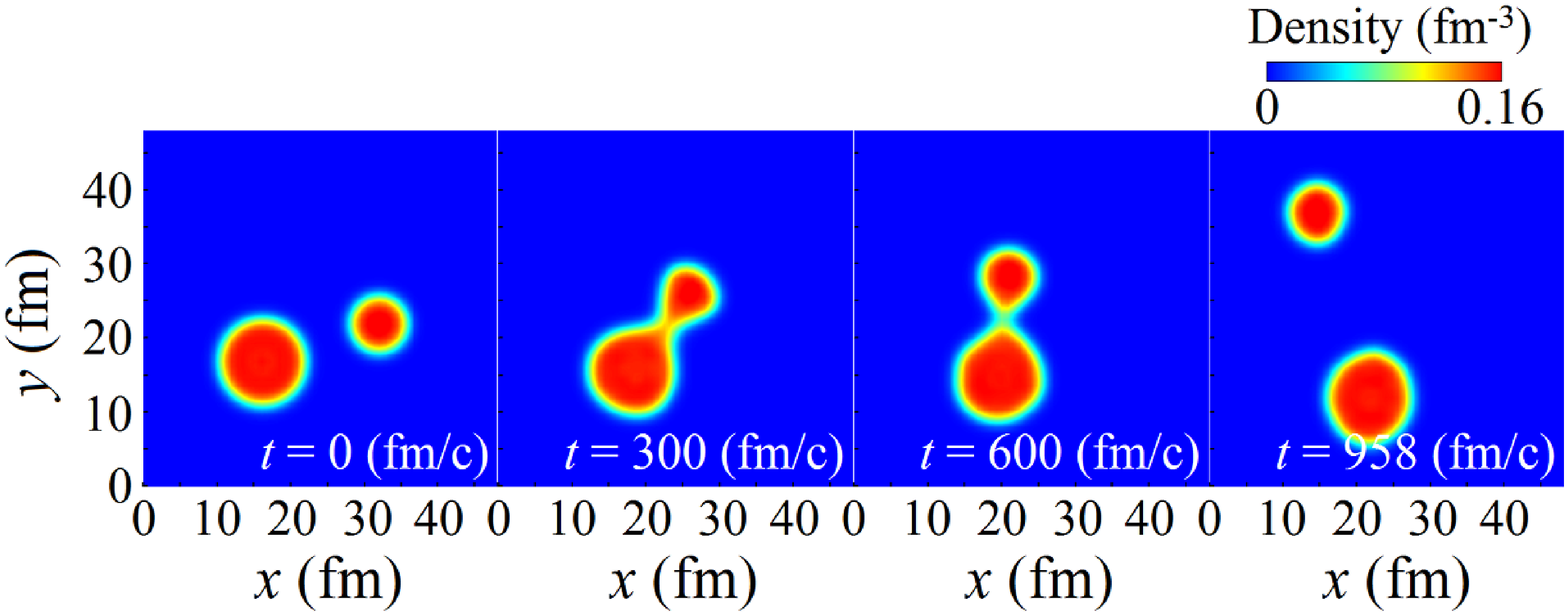}
   \end{center}\vspace{-3mm}
   \caption{(Color online)
   Snapshots of density distribution of the \RE{40}{Ca}+{124}{Sn} 
   reaction at $E_\mathrm{lab}=170$~MeV and $b=$ 3.96~fm, 
   just outside the fusion critical impact parameter.
   }
   \label{FIG:TDPLOT(40Ca+124Sn)}
\end{figure}
%&&&&&&&&&&&&&&&&&&&&&&&&&&&&&&&&&&&&&&&&&&&&&&&&&&

In Fig.~\ref{FIG:TDPLOT(40Ca+124Sn)}, we show snapshots of density 
distribution for the \RE{40}{Ca}+{124}{Sn} reaction at $b=3.96$~fm,
just outside the fusion critical impact parameter, $b_\mathrm{f}$. 
We find a formation of a neck between the projectile and the target 
during the collision. As will be shown later, several nucleons are exchanged 
between the projectile and the target at this impact parameter. We find 
a formation of the neck for the impact parameter region smaller than 
$b \sim$ 4.25~fm where the TKEL becomes appreciable.

%&&&&&&&&&&&&&&&&&&&&&&&&&&&&&&&&&&&&&&&&&&&&&&&&&&
\begin{figure*}[t]
   \begin{center}
   \includegraphics[height=16cm]{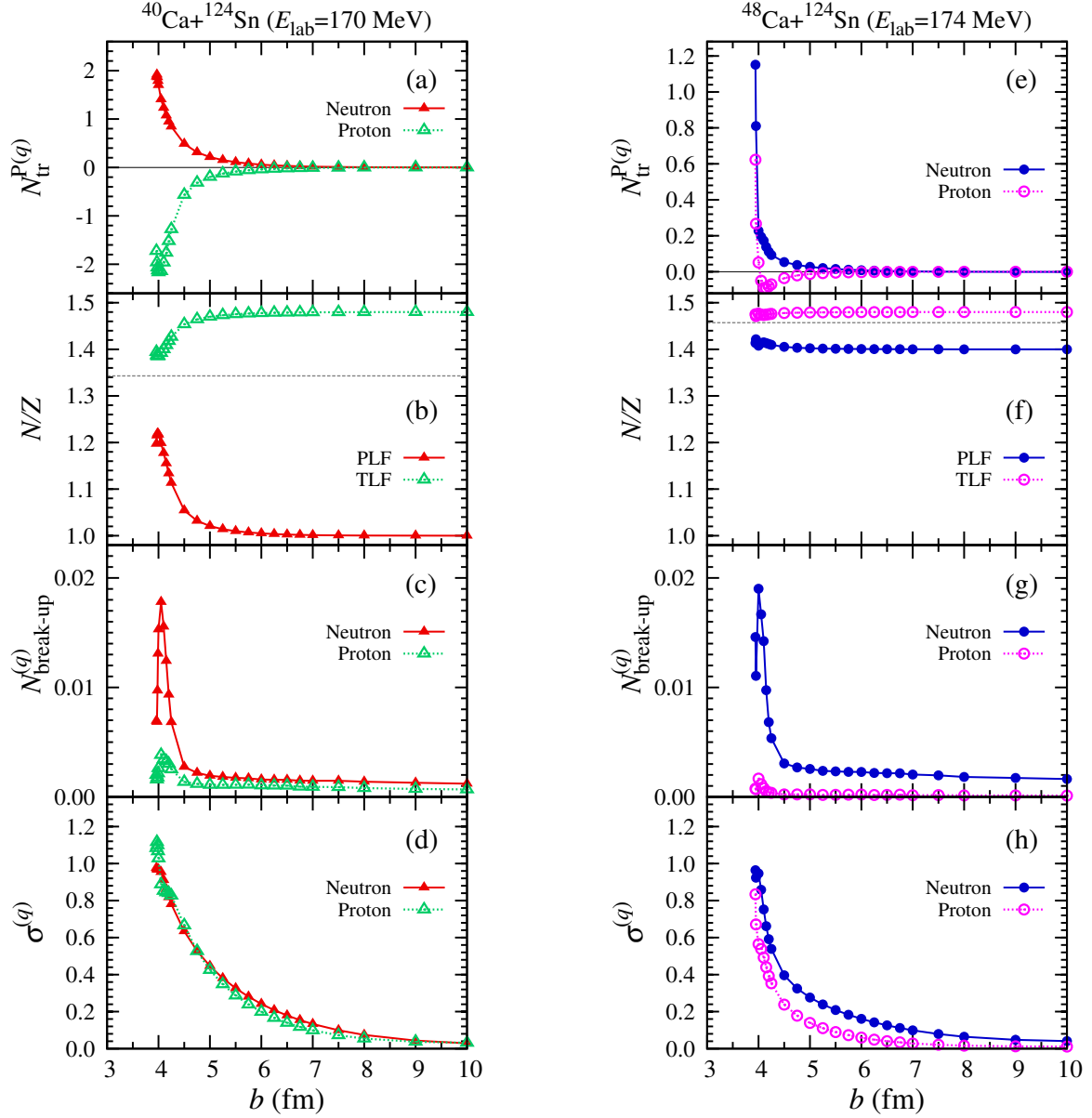}
   \end{center}\vspace{-3mm}
   \caption{(Color online)
   Left panels for \RE{40}{Ca}+{124}{Sn} at $E_\mathrm{lab}=$ 
   170~MeV and right panels for \RE{48}{Ca}+{124}{Sn} at 
   $E_\mathrm{lab}=$ 174~MeV.
   (a) and (e): Average number of transferred nucleons from the 
   target to the projectile. (b) and (f): Neutron-to-proton ratios, $N/Z$, 
   of the PLF and the TLF after collision. (c) and (g): Average number 
   of nucleons emitted to the continuum. (d) and (h): Fluctuation 
   of transferred nucleon number. The horizontal axis is the
   impact parameter $b$. In (b) and (f), the equilibrium $N/Z$ value 
   of the total system is indicated by a horizontal dashed line.
   }
   \label{FIG:Nave(40,48Ca+124Sn)}
\end{figure*}
%&&&&&&&&&&&&&&&&&&&&&&&&&&&&&&&&&&&&&&&&&&&&&&&&&&

We next consider the average number of transferred nucleons and 
its fluctuation. We denote the average number of nucleons in the PLF
as $N_\mathrm{PLF}^{(q)}$ ($q=n$ for neutrons, $p$ for protons), 
which is calculated from the density distribution at the final stage 
of the calculation,
\begin{equation}
N_\mathrm{PLF}^{(q)} = 
\int_\mathrm{around\,PLF} \hspace{-13mm}
d\bvecr\;\; \rho^{(q)}(\bvecr),
\end{equation}
where $\rho^{(q)}(\bvecr)$ is the density distribution of neutrons 
($q=n$) or protons ($q=p$). The spatial integration is achieved over 
a sphere whose center coincides with the center-of-mass of the PLF. 
The radius of the sphere is taken to be 10~fm. We calculate the average 
number of nucleons in the TLF in the same way taking the radius 
of 14~fm for the TLF. We summarize various expressions for the average 
number and the fluctuation of transferred nucleons in 
Appendix~\ref{App:average}.

We denote the neutron (proton) number of the projectile and
the target as $N_\mathrm{P}^{(q)}$ and $N_\mathrm{T}^{(q)}$, 
respectively. In general, there holds 
$N_\mathrm{PLF}^{(q)}+N_\mathrm{TLF}^{(q)} < 
N_\mathrm{P}^{(q)}+N_\mathrm{T}^{(q)}$, since some nucleons are 
emitted to the continuum by the breakup process. As will be shown later, 
however, the number of nucleons emitted to the continuum is very 
small in the present calculations. The average number of transferred 
nucleons from the projectile to the target, 
$N^{\mathrm{P}(q)}_\mathrm{tr}$, is given by
\begin{equation}
N^{\mathrm{P}(q)}_\mathrm{tr} = 
N_\mathrm{PLF}^{(q)}-N_\mathrm{P}^{(q)}.
\end{equation}

Figure~\ref{FIG:Nave(40,48Ca+124Sn)} shows the average number of
transferred nucleons, $N_\mathrm{tr}^{\mathrm{P}(q)}$, in (a) and 
(e), the neutron-to-proton ratios, $N/Z$, of the PLF and the TLF after 
collision in (b) and (f), the average number of nucleons emitted to the 
continuum in (c) and (g), and the fluctuation of the transferred nucleon 
number in (d) and (h), as functions of impact parameter $b$ for 
\RE{40,\,48}{Ca}+{124}{Sn} reactions. 

In Fig.~\ref{FIG:Nave(40,48Ca+124Sn)}~(a) and (e), the average 
number of transferred neutrons is shown by filled symbols connected 
with solid lines, while the average number of transferred protons is 
shown by open symbols connected with dotted lines. Positive values
indicate the increase of the projectile nucleons (transfer from $^{124}$Sn 
to $^{40,\,48}$Ca) and negative values indicate the decrease (transfer 
from $^{40,\,48}$Ca to $^{124}$Sn). As seen from 
Fig.~\ref{FIG:Nave(40,48Ca+124Sn)}~(a) and (e), a large value of 
average number of transferred nucleons is seen for 
\RE{40}{Ca}+{124}{Sn} at the impact parameter region close to 
the fusion critical impact parameter, while the average number of 
transferred nucleons is small for \RE{48}{Ca}+{124}{Sn}. 

We show in Fig.~\ref{FIG:Nave(40,48Ca+124Sn)}~(b) and (f) the 
neutron-to-proton ratios, $N/Z$, of the PLF and the TLF. For the PLF, it is 
given by $N_\mathrm{PLF}^{(n)}/N_\mathrm{PLF}^{(p)}$, and for 
the TLF by $N_\mathrm{TLF}^{(n)}/N_\mathrm{TLF}^{(p)}$. 
Before the collision, the $N/Z$ ratio is given by 1.00 for $^{40}$Ca, 
1.40 for $^{48}$Ca, and 1.48 for $^{124}$Sn.
The $N/Z$ ratio of the PLF (TLF) is denoted by filled (open) symbols 
connected with solid (dotted) lines. We also denote the $N/Z$ ratio 
of the total system by a horizontal dashed line in the figure, 1.34 for 
\RE{40}{Ca}+{124}{Sn} and 1.46 for \RE{48}{Ca}+{124}{Sn}. 
We find the nucleons are transferred towards the direction of the charge 
equilibrium. Namely, protons are transferred from $^{40}$Ca to 
$^{124}$Sn, while neutrons are transferred from $^{124}$Sn to 
$^{40}$Ca in the \RE{40}{Ca}+{124}{Sn} reaction. The $N/Z$ 
ratios of the projectile and the target do not differ much for 
\RE{48}{Ca}+{124}{Sn}, and we find a small number of transferred 
nucleons on average for this reaction. The average number of 
transferred nucleons decreases rapidly as the impact parameter 
increases. For the impact parameter region larger than $b \sim 6$~fm, 
the average number of transferred nucleons almost vanishes.

In Fig.~\ref{FIG:Nave(40,48Ca+124Sn)}~(c) and (g), we show the 
average number of nucleons emitted to the continuum, 
$N_\mathrm{break\mathchar`-up}^{(q)}
\equiv$ $(N_\mathrm{P}^{(q)}+N_\mathrm{T}^{(q)})-
(N_\mathrm{PLF}^{(q)}+N_\mathrm{TLF}^{(q)})$, 
during the time evolution. The average number of neutrons 
(protons) emitted to the continuum is denoted by filled (open) 
symbols connected with solid (dotted) lines. As seen in the figure, 
the number of emitted nucleons is very small. The maximum value, 
about 0.02, is seen at the impact parameter close to the fusion 
critical impact parameter.

In Fig.~\ref{FIG:Nave(40,48Ca+124Sn)}~(d) and (h), we show the 
fluctuation of the transferred nucleon number. The expression for the 
fluctuation is given by Eq.~(\ref{fluctuation_sigma}). The fluctuation 
of the transferred neutron (proton) number is denoted by filled (open) 
symbols connected with solid (dotted) lines. We find the fluctuation 
decreases as the impact parameter increases. The fluctuation decreases 
more slowly than the average number of transferred nucleons as a 
function of impact parameter. We also find the fluctuation of 
\RE{48}{Ca}+{124}{Sn} is somewhat smaller than but comparable in 
magnitude to that of \RE{40}{Ca}+{124}{Sn}, although the average 
number is vanishingly small for \RE{48}{Ca}+{124}{Sn}.

%&&&&&&&&&&&&&&&&&&&&&&&&&&&&&&&&&&&&&&&&&&&&&&&&&&
\begin{figure}[b]
   \begin{center}
   \includegraphics[width=8.6cm]{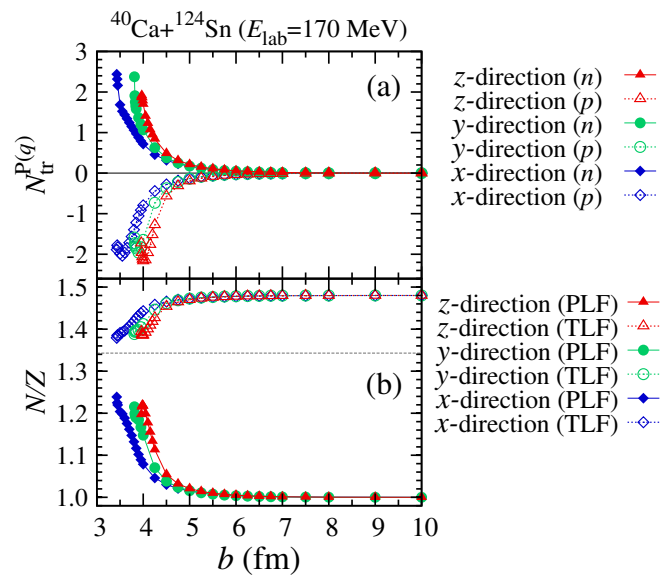}
   \end{center}\vspace{-3mm}
   \caption{(Color online)
   Comparison of calculated results for different initial
   orientations of $^{124}$Sn in the \RE{40}{Ca}+{124}{Sn} 
   reaction at $E_\mathrm{lab}=$ 170~MeV.
   (a): Average number of transferred nucleons from the 
   target to the projectile. (b): Neutron-to-proton ratios, $N/Z$, 
   of the PLF and the TLF after collision. The horizontal axis is the
   impact parameter $b$. The initial orientations of $^{124}$Sn are
   indicated in legends.
   }
   \label{FIG:Nave(40Ca+124Sn)XYZ}
\end{figure}
%&&&&&&&&&&&&&&&&&&&&&&&&&&&&&&&&&&&&&&&&&&&&&&&&&&

As mentioned in the beginning of this section, we placed
the $^{124}$Sn nucleus which is oblately deformed with 
$\beta \sim 0.11$ so that the symmetry axis is perpendicular 
to the reaction plane in the initial configuration. 
Namely, the symmetry axis of $^{124}$Sn is set parallel to 
the $z$-axis. To take fully account of the deformation effect, 
we should achieve an average over initial orientations of the 
$^{124}$Sn. However, since calculations of a number of initial 
orientations require large computational costs, we do not 
achieve the orientational average but show results of a 
specific initial orientation in the present paper. We here 
briefly discuss the difference of the reaction dynamics 
depending on the initial orientations.

In Fig.~\ref{FIG:Nave(40Ca+124Sn)XYZ}, we show the average 
number of transferred nucleons in (a) and the neutron-to-proton 
ratios, $N/Z$, of the PLF and the TLF after collision in (b), 
for three cases of different initial orientations of $^{124}$Sn 
in the \RE{40}{Ca}+{124}{Sn} reaction. Red triangles are the 
same results as those shown in Fig.~\ref{FIG:Nave(40,48Ca+124Sn)}~(a) 
and (b) where the symmetry axis of $^{124}$Sn is chosen parallel to 
the $z$-axis. Green circles correspond to the cases of the symmetry 
axis set parallel to the $y$-axis (the direction of impact 
parameter vector). Blue diamonds correspond to the cases of the 
symmetry axis set parallel to the $x$-axis (the incident direction).

From the figure, we find a rather small difference among three
cases of different initial orientations of $^{124}$Sn.
The prominent difference appears only at small impact parameter
region. It comes from the difference of the fusion critical impact 
parameters. Since $^{124}$Sn is oblately deformed, the
Coulomb barrier height is the largest when the symmetry
axis of the $^{124}$Sn is parallel to the $x$-axis (the incident 
direction).

\subsubsection{Transfer probabilities}

%&&&&&&&&&&&&&&&&&&&&&&&&&&&&&&&&&&&&&&&&&&&&&&&&&&
\begin{figure}[b]
   \begin{center}
   \includegraphics[width=8.6cm]{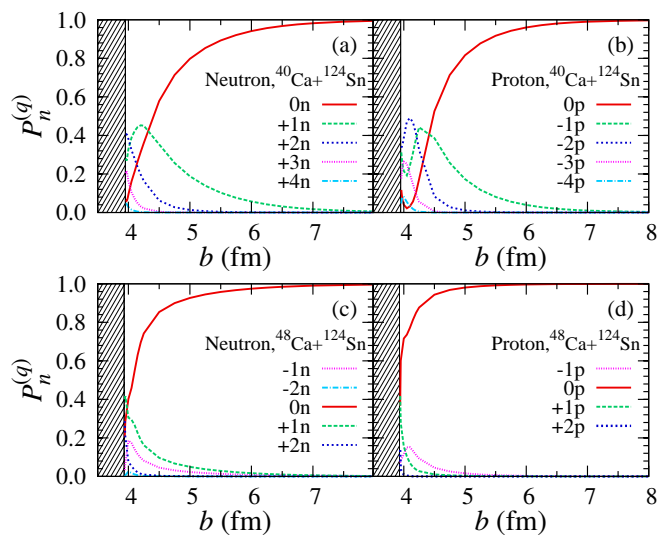}
   \end{center}\vspace{-3mm}
   \caption{(Color online)
   Neutron (left panels) and proton (right panels) transfer 
   probabilities as functions of impact parameter $b$. (a) and (b): 
   Results for the reactions of \RE{40}{Ca}+{124}{Sn} at $E_\mathrm{lab}=$ 
   170~MeV. (c) and (d): Results for the reactions of \RE{48}{Ca}+{124}{Sn} 
   at $E_\mathrm{lab}=$ 174~MeV. The positive (negative) number of 
   transferred nucleons represents the number of nucleons added to 
   (removed from) the projectile. Shaded regions at small impact parameter 
   ($b \le 3.95$~fm for \RE{40}{Ca}+{124}{Sn} and $b \le 3.93$~fm 
   for \RE{48}{Ca}+{124}{Sn}) correspond to the fusion reactions.
   }
   \label{FIG:Pn(40,48Ca+124Sn)}
\end{figure}
%&&&&&&&&&&&&&&&&&&&&&&&&&&&&&&&&&&&&&&&&&&&&&&&&&&

%&&&&&&&&&&&&&&&&&&&&&&&&&&&&&&&&&&&&&&&&&&&&&&&&&&
\begin{figure}[t]
   \begin{center}
   \includegraphics[width=8.6cm]{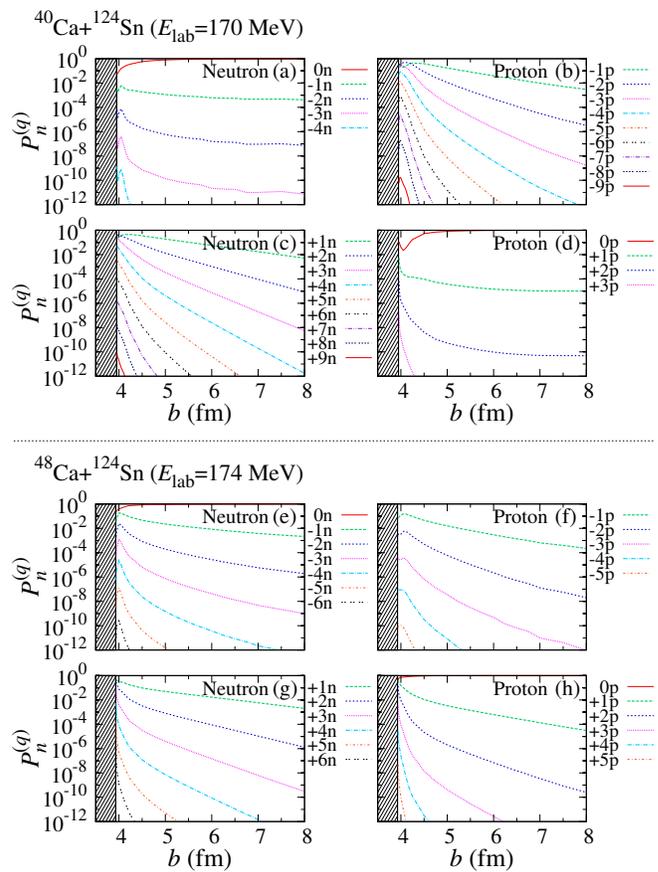}
   \end{center}\vspace{-3mm}
   \caption{(Color online)
   Transfer probabilities in Fig.~\ref{FIG:Pn(40,48Ca+124Sn)}
   are shown in logarithmic scale. Nucleon transfer probabilities 
   opposite to the direction of the charge equilibrium, which are
   not included in Fig.~\ref{FIG:Pn(40,48Ca+124Sn)}, are shown
   as well.
   }
   \label{FIG:Pn(40,48Ca+124Sn)_log}
\end{figure}
%&&&&&&&&&&&&&&&&&&&&&&&&&&&&&&&&&&&&&&&&&&&&&&&&&&

We next show transfer probabilities as functions of impact parameter
which are obtained from the final wave functions using the number 
projection procedure of Eq.~(\ref{P_n_projection}).
The nucleon transfer probabilities, $P_n^{(q)}(b)$, are shown in 
Fig.~\ref{FIG:Pn(40,48Ca+124Sn)} (linear scale) and in 
Fig.~\ref{FIG:Pn(40,48Ca+124Sn)_log} (logarithmic scale). 
Top panels of Fig.~\ref{FIG:Pn(40,48Ca+124Sn)} ((a) and (b)) 
and top panels of Fig.~\ref{FIG:Pn(40,48Ca+124Sn)_log} 
((a), (b), (c), and (d)) show results of the \RE{40}{Ca}+{124}{Sn} 
reaction, while lower panels of Fig.~\ref{FIG:Pn(40,48Ca+124Sn)} 
((c) and (d)) and lower panels of Fig.~\ref{FIG:Pn(40,48Ca+124Sn)_log} 
((e), (f), (g), and (h)) show results of the \RE{48}{Ca}+{124}{Sn} 
reaction. In these figures, shaded regions at small impact parameter
($b \le 3.95$~fm for \RE{40}{Ca}+{124}{Sn} and $b \le 3.93$~fm 
for \RE{48}{Ca}+{124}{Sn}) correspond to the fusion reactions. 
The positive (negative) number of transferred nucleons represents 
the number of nucleons added to (removed from) the projectile. 

From the figure, we find that probabilities of single-nucleon transfer 
(green dashed lines) extend to a large impact parameter region. 
As the number of transferred nucleons increases, the reaction 
probability is sizable only at a small impact parameter region, close to 
the fusion critical impact parameter. 

The directions of the transfer processes are the same as those 
we observed in the average number of transferred nucleons in 
Fig.~\ref{FIG:Nave(40,48Ca+124Sn)}~(a) and (e). Namely, 
in the case of \RE{40}{Ca}+{124}{Sn} 
(Fig.~\ref{FIG:Pn(40,48Ca+124Sn)}~(a) and (b)), protons are 
transferred from $^{40}$Ca to $^{124}$Sn and neutrons are 
transferred from $^{124}$Sn to $^{40}$Ca, the directions towards 
the charge equilibrium. We note that the transfer probabilities towards 
the opposite directions, proton transfer from $^{124}$Sn to 
$^{40}$Ca and neutron transfer from $^{40}$Ca to 
$^{124}$Sn, are very small and are hardly seen in the linear 
scale figure (Fig.~\ref{FIG:Pn(40,48Ca+124Sn)}~(a) and (b)). 
In the logarithmic scale 
(Fig.~\ref{FIG:Pn(40,48Ca+124Sn)_log}~(a), (b), (c), and (d)), 
we find the transfer probabilities towards the opposite direction to 
the charge equilibrium are smaller than those towards the charge 
equilibrium by at least an order of magnitude. In the case of 
\RE{48}{Ca}+{124}{Sn} reaction 
(Fig.~\ref{FIG:Pn(40,48Ca+124Sn)}~(c) and (d), 
Fig.~\ref{FIG:Pn(40,48Ca+124Sn)_log}~(e), (f), (g), and (h)), 
the transfer probabilities towards both directions are the same order 
of magnitude. This is consistent with the fact that the average number 
of transferred nucleons is very small as shown in 
Fig.~\ref{FIG:Nave(40,48Ca+124Sn)}~(e).

\subsubsection{Transfer cross sections}

%&&&&&&&&&&&&&&&&&&&&&&&&&&&&&&&&&&&&&&&&&&&&&&&&&&
\begin{figure}[t]
   \begin{center}
   \includegraphics[width=8.6cm]{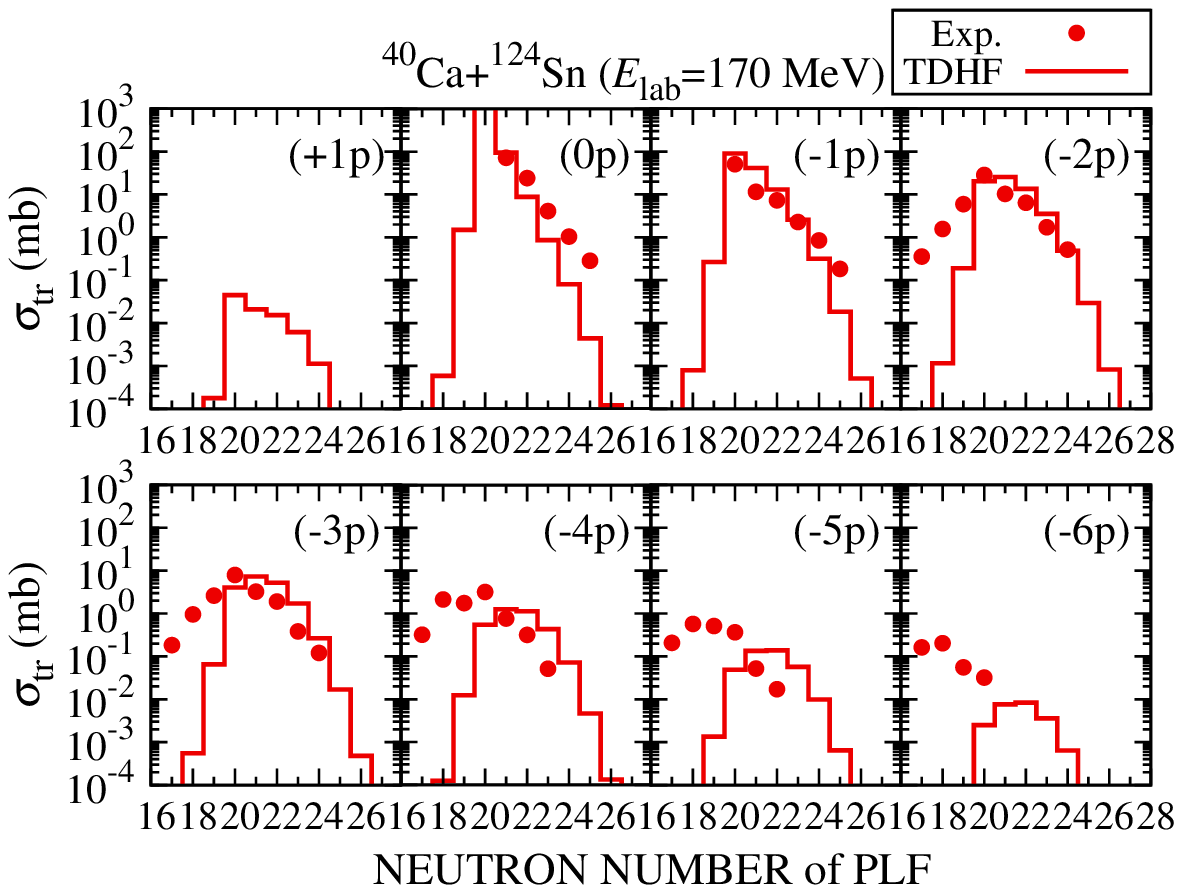}
   \end{center}\vspace{-3mm}
   \caption{(Color online)
   Cross sections for transfer channels classified according to
   the change of the proton number of the PLF from $^{40}$Ca, as 
   functions of neutron number of the PLF for the 
   \RE{40}{Ca}+{124}{Sn} reaction at $E_\mathrm{lab}=$ 170~MeV. 
   Red filled circles denote measured cross sections and red solid lines 
   denote results of the TDHF calculations. The number of transferred 
   protons is indicated as ($x$p) ($-6 \le x \le +1$). The measured 
   cross sections have been reported in Ref.~\cite{Corradi(40Ca+124Sn)}.
   }
   \label{FIG:NTCS(40Ca+124Sn)}
\end{figure}
%&&&&&&&&&&&&&&&&&&&&&&&&&&&&&&&&&&&&&&&&&&&&&&&&&&

\begin{figure}[t]
   \begin{center}
   \includegraphics[width=8.6cm]{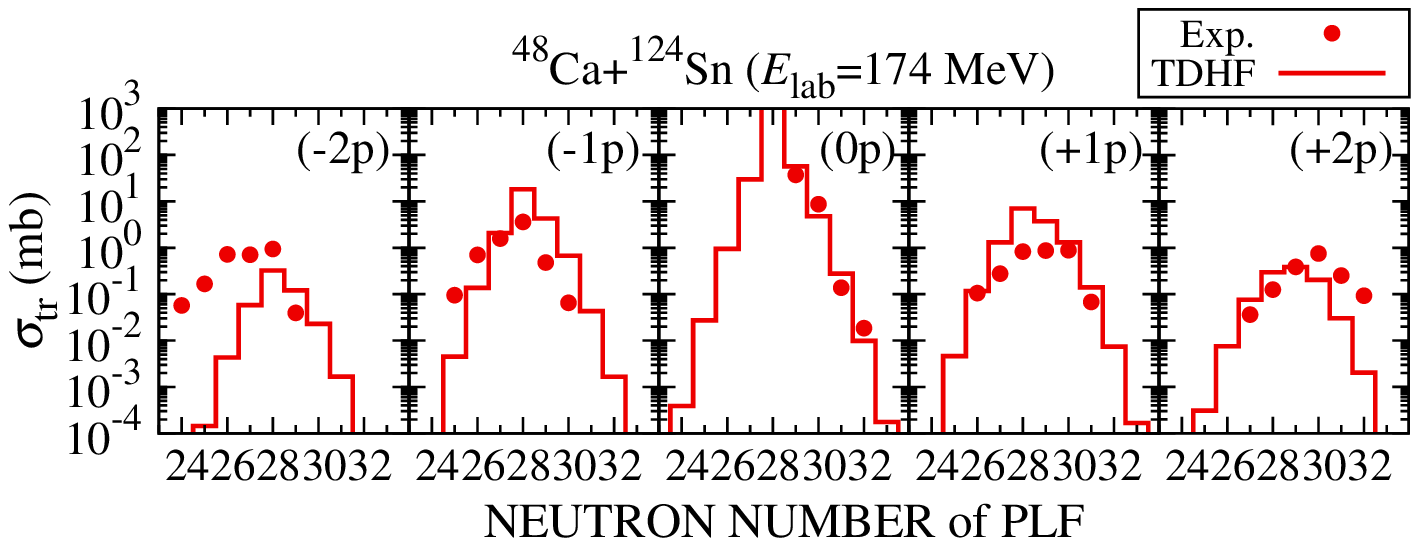}
   \end{center}\vspace{-3mm}
   \caption{(Color online)
   Cross sections for transfer channels classified according to
   the change of the proton number of the PLF from $^{48}$Ca, as 
   functions of neutron number of the PLF for the 
   \RE{48}{Ca}+{124}{Sn} reaction at $E_\mathrm{lab}=$ 174~MeV. 
   Red filled circles denote measured cross sections and red solid lines 
   denote results of the TDHF calculations. The number of transferred 
   protons is indicated as ($x$p) ($-2 \le x \le +2$). The measured 
   cross sections have been reported in Ref.~\cite{Corradi(48Ca+124Sn)}.
   }
   \label{FIG:NTCS(48Ca+124Sn)}
\end{figure}

Integrating the transfer probabilities over impact parameter,
we obtain transfer cross sections. The results are shown in 
Fig.~\ref{FIG:NTCS(40Ca+124Sn)} for \RE{40}{Ca}+{124}{Sn} 
and in Fig.~\ref{FIG:NTCS(48Ca+124Sn)} for \RE{48}{Ca}+{124}{Sn}.

We first examine the \RE{40}{Ca}+{124}{Sn} reaction.
Figure~\ref{FIG:NTCS(40Ca+124Sn)} shows the transfer cross sections 
classified according to the change of the proton number of the PLF from 
$^{40}$Ca, as functions of neutron number of the PLF. Red filled 
circles denote measured cross sections and red solid lines denote results 
of the TDHF calculations. We show transfer cross sections of one proton 
added to ($+1$p) through six proton removed from ($-6$p) $^{40}$Ca.

We find the experimental data are reasonably reproduced by the TDHF 
calculations for cross sections without proton transfer shown in ($0$p) 
panel, although the cross sections are somewhat underestimated as
the number of transferred neutrons increases.
Zero- to four-neutron pick-up channels shown in ($-1$p), ($-2$p), and 
($-3$p) panels are also reproduced reasonably.
The calculated cross sections towards the direction opposite to 
the charge equilibrium are small, consistent with the observation 
in transfer probabilities shown in 
Fig.~\ref{FIG:Pn(40,48Ca+124Sn)_log}~(a) and (d).

As the number of transferred protons increases,
there appear some discrepancies between the TDHF calculations 
and the measurements. When more than one protons are transferred,
the TDHF calculation underestimates measured cross sections of
neutron removal channels ($N<20$).
For five- and six-proton removal channels, ($-5$p) and ($-6$p), 
the TDHF cross sections become too small compared with the 
measurements. We also find a shift of the peak position towards 
the larger neutron number.

In Ref.~\cite{Corradi(40Ca+124Sn)}, cross sections calculated by 
the GRAZING code \cite{GRAZING} were compared with the 
measurements. In the GRAZING calculation, a similar discrepancy 
was observed. As the origin of the discrepancy, the significance of 
the evaporation effects has been mentioned \cite{Corradi(40Ca+124Sn)}. 
We will compare our results with those of the GRAZING calculations 
in Sec.~\ref{Sec:Comparison}.

We note that particle evaporation processes are not taken into
account sufficiently in the present calculation.
In Fig.~\ref{FIG:Theta+TKEL(40,48Ca+124Sn)}~(b), 
we find the TKEL of as large as 25~MeV at a small impact parameter 
region where appreciable multinucleon transfer probabilities are found. 
The amount of the TKEL is sufficiently large to emit some 
nucleons to the continuum. However, as we saw in 
Fig.~\ref{FIG:Nave(40,48Ca+124Sn)}~(c), the average number of 
nucleons emitted to the continuum is very small, the maximum value 
is only 0.02. Although we have not yet estimated the number of 
evaporated nucleons, the inclusion of the evaporation processes is 
expected to reduce the discrepancy as follows. Neutron evaporation 
processes will shift the peak position of the transfer cross sections 
towards the smaller neutron number (left direction in 
Fig.~\ref{FIG:NTCS(40Ca+124Sn)}). We may also expect that 
proton evaporation processes will shift cross section of $n$-proton 
removal channels to ($n+1$)-proton removal channels.

%&&&&&&&&&&&&&&&&&&&&&&&&&&&&&&&&&&&&&&&&&&&&&&&&&&
%&&&&&&&&&&&&&&&&&&&&&&&&&&&&&&&&&&&&&&&&&&&&&&&&&&

Figure~\ref{FIG:NTCS(48Ca+124Sn)} shows transfer cross sections of 
\RE{48}{Ca}+{124}{Sn} reaction. The cross sections obtained 
from the TDHF calculations are in good agreement with the experimental 
data for zero- and one-proton transfer channels, ($0$p) and ($\pm1$p). 
For two-proton transfer channels ($\pm 2$p), however, our TDHF 
calculations underestimate the cross sections. In the case of two-proton 
removal channels ($-2$p), the peak position shifts towards larger neutron 
number, while in the case of two-proton pickup channels ($+2$p), the 
peak position shifts towards smaller neutron number. The underestimation 
in the ($-2$p) channels may be remedied by taking into account the 
neutron evaporation processes as in the case of \RE{40}{Ca}+{124}{Sn} 
reaction. However, the underestimation in the ($+2$p) channels may not. 
A similar discrepancy was reported in the GRAZING calculation 
\cite{Corradi(48Ca+124Sn)}. In Ref.~\cite{Corradi(48Ca+124Sn)}, 
more complex mechanisms such as neutron-proton pair transfer and/or 
$\alpha$-cluster transfer have been advocated for the origin of the 
discrepancy.

\subsection{\RE{40}{Ca}+{208}{Pb} reaction}

In this subsection, we present results for the reactions of 
\RE{40}{Ca}+{208}{Pb} at $E_\mathrm{lab}=$ 235 and 249~MeV
($E_\mathrm{c.m.}\simeq$ 197.1 and 208.8~MeV), for which 
measurements have been reported in Ref.~\cite{Szilner(40Ca+208Pb)2}.
This system has $Z_\mathrm{P}Z_\mathrm{T}=$ 1640, close to 1600.
Therefore, we expect an appearance of the indication of the 
fusion-hindrance. We estimate the Coulomb barrier 
height of this system using the frozen-density approximation, giving 
$V_\mathrm{B}\approx$ 178.4~MeV. Since the collision energies 
are higher than the barrier height, we find finite values of the fusion 
critical impact parameter $b_\mathrm{f}$, as in the 
\RE{40,\,48}{Ca}+{124}{Sn} reactions. They are given by 
$b_\mathrm{f}=3.81$ and 4.55~fm at $E_\mathrm{lab}=$ 
235 and 249~MeV, respectively.

\subsubsection{Overview of the reactions}

\begin{figure}[t]
   \begin{center}
   \includegraphics[width=6.5cm]{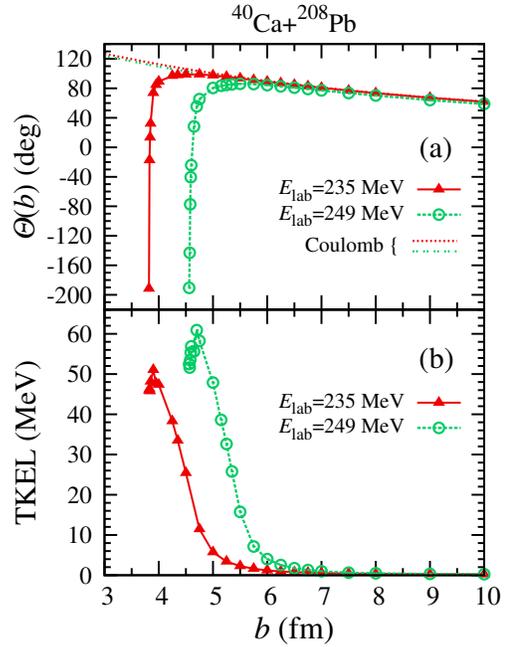}
   \end{center}\vspace{-3mm}
   \caption{(Color online)
   Deflection function (a) and total kinetic energy loss (b) as functions 
   of impact parameter $b$ for the reactions of \RE{40}{Ca}+{208}{Pb} 
   at $E_\mathrm{lab}=$ 235 and 249~MeV. Results for the reactions 
   at $E_\mathrm{lab}=$ 235~MeV are denoted by red filled triangles 
   connected with solid lines, while results for the reactions at 
   $E_\mathrm{lab}=$ 249~MeV are denoted by green open circles 
   connected with dashed lines. In (a), we also show deflection functions 
   for the pure Coulomb trajectories at $E_\mathrm{lab}=$ 235~MeV
   by a red dotted line and at $E_\mathrm{lab}=$ 249~MeV
   by a green two-dot chain line.
   }
   \label{FIG:Theta+TKEL(40Ca+208Pb)}
\end{figure}

We first present an overview of the reaction dynamics.
In Fig.~\ref{FIG:Theta+TKEL(40Ca+208Pb)}, we show 
the deflection function in (a) and the TKEL in (b), as functions of 
impact parameter. Results for the reaction at $E_\mathrm{lab}=$ 235 
MeV are denoted by red filled triangles connected with solid lines, while 
results for the reaction at $E_\mathrm{lab}=$ 249~MeV are denoted 
by green open circles connected with dotted lines. 
In Fig.~\ref{FIG:Theta+TKEL(40Ca+208Pb)}~(a), we also show 
deflection functions for the pure Coulomb trajectories at $E_\mathrm{lab}=$
235~MeV by a red dotted line and at $E_\mathrm{lab}=$ 249~MeV 
by a green two-dot chain line. In this system, we find an increase of 
the TKEL up to around 50~MeV and 60~MeV for the incident energies 
of 235~MeV and 249~MeV, respectively. This maximum value of 
TKEL is about a factor of two larger than the case of 
\RE{40,\,48}{Ca}+{124}{Sn} reactions. We find the difference 
of the TKEL between these systems, \RE{40}{Ca}+{208}{Pb} and 
\RE{40,\,48}{Ca}+{124}{Sn}, comes from properties of the neck 
whose formation is observed when the TKEL becomes substantial. 

%&&&&&&&&&&&&&&&&&&&&&&&&&&&&&&&&&&&&&&&&&&&&&&&&&&
%&&&&&&&&&&&&&&&&&&&&&&&&&&&&&&&&&&&&&&&&&&&&&&&&&&

%&&&&&&&&&&&&&&&&&&&&&&&&&&&&&&&&&&&&&&&&&&&&&&&&&&
\begin{figure}[b]
   \begin{center}
   \includegraphics[width=8.6cm]{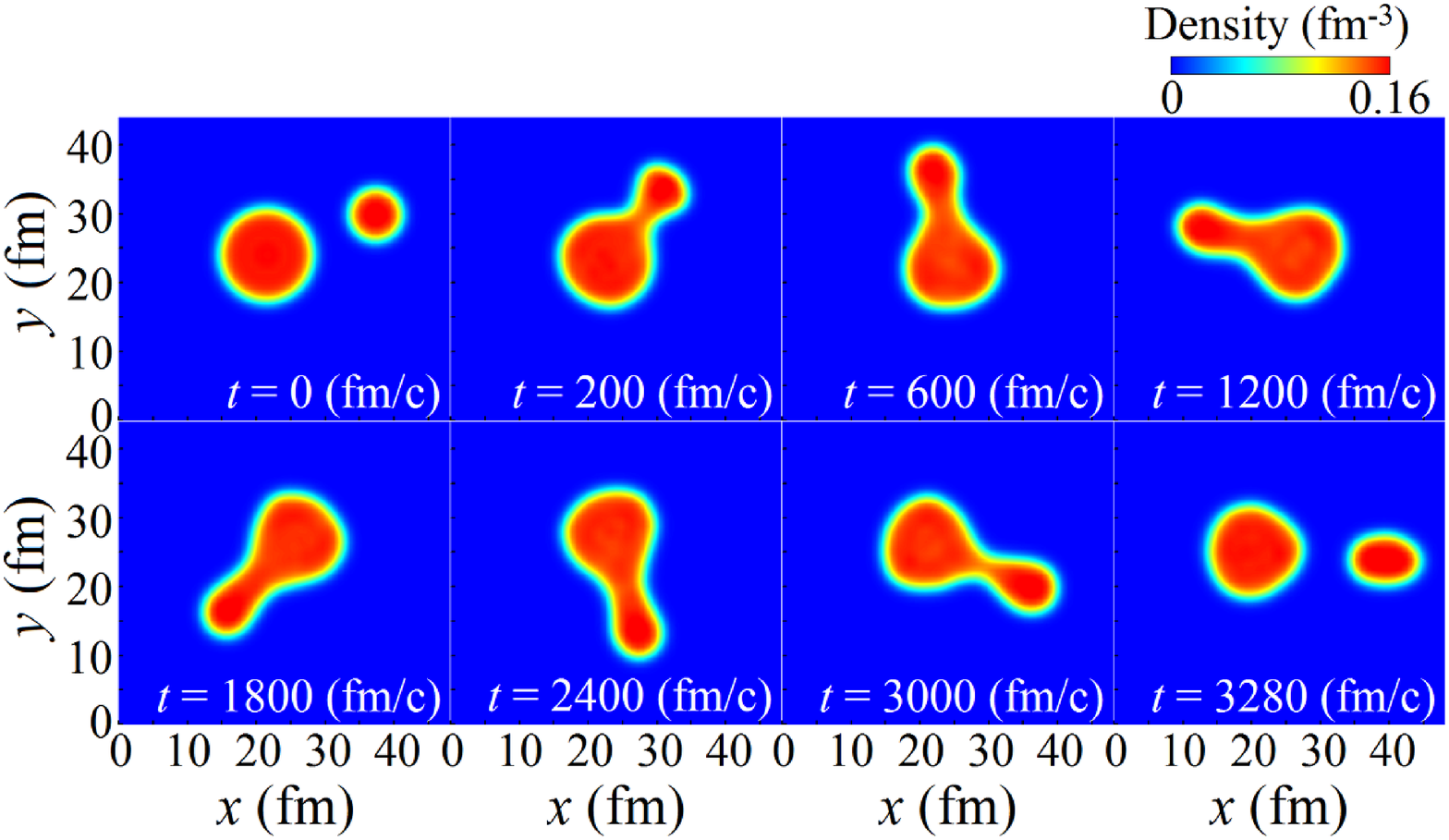}
   \end{center}\vspace{-3mm}
   \caption{(Color online)
   Snapshots of density distribution of the \RE{40}{Ca}+{208}{Pb} 
   reaction at $E_\mathrm{lab}=$ 249~MeV and $b=$ 4.56~fm, 
   just outside the fusion critical impact parameter.
   }
   \label{FIG:TDPLOT(40Ca+208PbE249,b4.56)}
\end{figure}
%&&&&&&&&&&&&&&&&&&&&&&&&&&&&&&&&&&&&&&&&&&&&&&&&&&

In Fig.~\ref{FIG:TDPLOT(40Ca+208PbE249,b4.56)}, we show 
snapshots of density distribution for the \RE{40}{Ca}+{208}{Pb} 
reaction at $E_\mathrm{lab}=$ 249~MeV and $b=4.56$~fm, just 
outside the fusion critical impact parameter. The neck is seen to be 
formed solidly for a long period from $t=$ 200~fm/c to 3000~fm/c. 
This process may be regarded as a quasi-fission. As will be shown below, 
a number of nucleons are transferred from $^{208}$Pb to $^{40}$Ca 
at this impact parameter. 

%&&&&&&&&&&&&&&&&&&&&&&&&&&&&&&&&&&&&&&&&&&&&&&&&&&
\begin{figure}[t]
   \begin{center}
   \includegraphics[width=8cm]{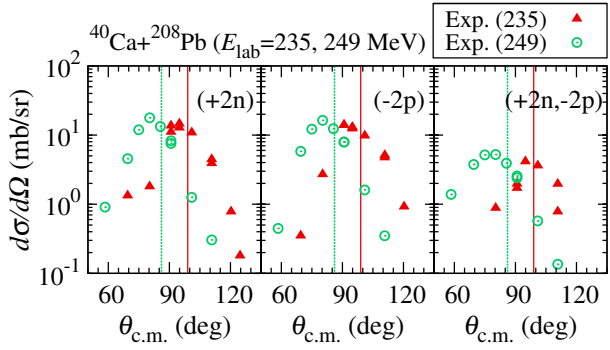}
   \end{center}\vspace{-3mm}
   \caption{(Color online)
   Differential cross sections of representative transfer 
   channels as functions of scattering angle in the center-of-mass 
   frame for the \RE{40}{Ca}+{208}{Pb} reactions at 
   $E_\mathrm{lab}=$ 235 and 249~MeV.
   The Coulomb rainbow angle obtained from the TDHF trajectories 
   is denoted by red solid (green dotted) vertical lines for 
   $E_\mathrm{lab}=$ 235 (249) MeV. They are compared with
   measured differential cross sections, red filled triangles 
   (green open circles) for $E_\mathrm{lab}=$ 235 (249) MeV, 
   which have been reported in Ref.~\cite{Szilner(40Ca+208Pb)2}.
    }
   \label{FIG:DCS(40Ca+208Pb)_exp}
\end{figure}
%&&&&&&&&&&&&&&&&&&&&&&&&&&&&&&&&&&&&&&&&&&&&&&&&&&

The period of the neck formation is longer in the 
present \RE{40}{Ca}+{208}{Pb} case than that in the 
\RE{40,\,48}{Ca}+{124}{Sn} cases. We find the neck formation for 
the periods of 1000-3000~fm/c and more for the present system 
depending on the impact parameter, while it is at most 300~fm/c in the 
\RE{40,\,48}{Ca}+{124}{Sn} systems. We consider this difference 
is related to the different $Z_\mathrm{P}Z_\mathrm{T}$ values of 
these systems. Since $Z_\mathrm{P}Z_\mathrm{T}\gtrsim$ 1600 
in the present system, fusion reactions are hindered by the quasi-fission 
process. Namely, there appears a certain impact parameter region 
in which binary final fragments are produced after a rather solid 
neck formation during the collision.

The Coulomb rainbow angle is $\theta_\mathrm{r}\simeq$ 
99$^\circ$ for the reaction at $E_\mathrm{lab}=$ 235~MeV 
and $\theta_\mathrm{r}\simeq$ 86$^\circ$ for the reaction at 
$E_\mathrm{lab}=$ 249~MeV, respectively. The deflection 
function becomes negative at the small impact parameter region, 
reaching $-200^\circ$ just outside $b_\mathrm{f}$. 
In Fig.~\ref{FIG:DCS(40Ca+208Pb)_exp}, we compare the 
Coulomb rainbow angles for the \RE{40}{Ca}+{208}{Pb} 
reactions at $E_\mathrm{lab}=$ 235 and 249~MeV with measured 
differential cross sections which have been reported in 
Ref.~\cite{Szilner(40Ca+208Pb)2}. Red filled triangles denote 
measured cross sections for $E_\mathrm{lab}=$ 235~MeV, 
while green open circles denote those for $E_\mathrm{lab}=$ 
249~MeV. The Coulomb rainbow angle obtained from the TDHF 
trajectories is denoted by red solid (green dotted) vertical lines 
for $E_\mathrm{lab}=$ 235 (249) MeV. We find the peak 
positions of measured angular distributions are reasonably 
reproduced by the TDHF calculation.

%&&&&&&&&&&&&&&&&&&&&&&&&&&&&&&&&&&&&&&&&&&&&&&&&&&
\begin{figure}[t]
   \begin{center}
   \includegraphics[height=16cm]{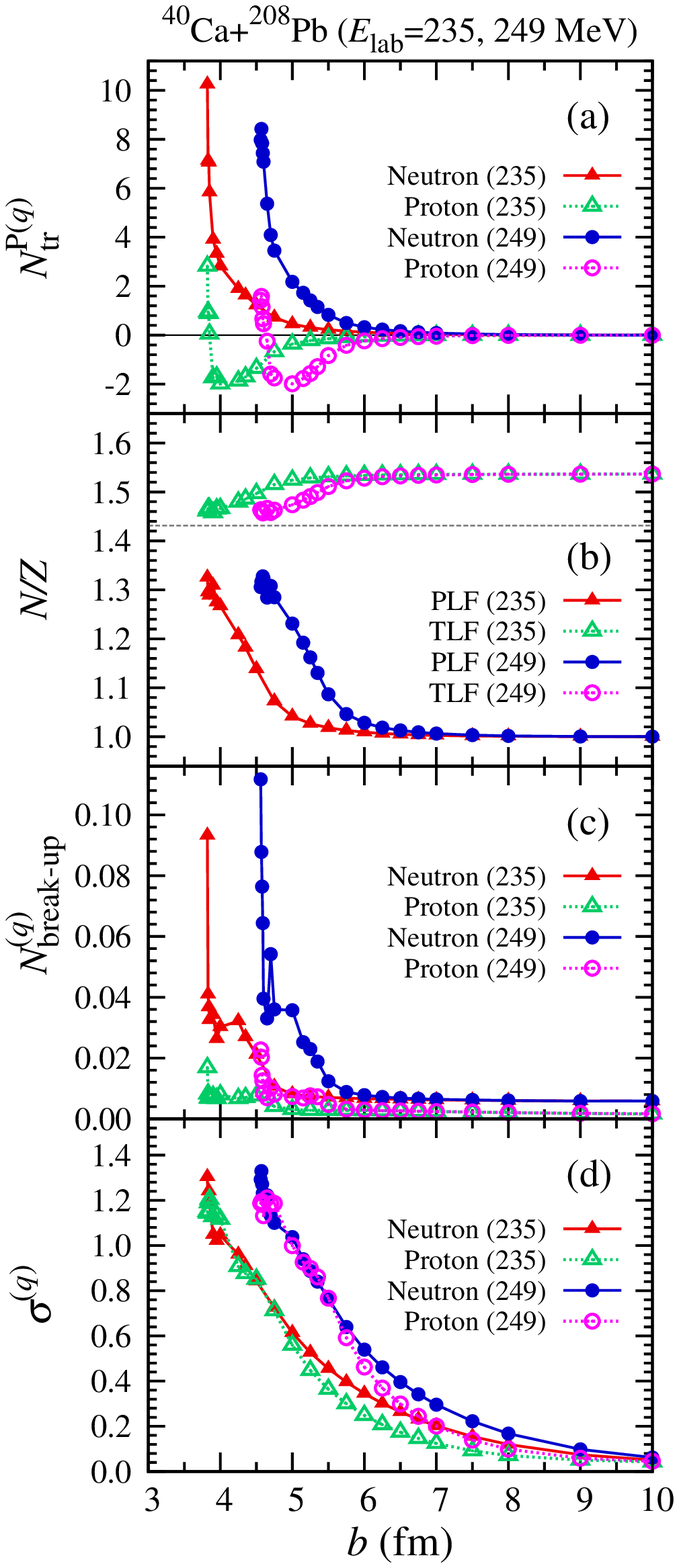}
   \end{center}\vspace{-4mm}
   \caption{(Color online)
   The \RE{40}{Ca}+{208}{Pb} reactions at $E_\mathrm{lab}=$ 
   235 and 249~MeV. (a): Average number of transferred nucleons 
   from the target to the projectile. (b): Neutron-to-proton ratios, $N/Z$, 
   of the PLF and the TLF after collision. (c): Average number 
   of nucleons emitted to the continuum. (d): Fluctuation 
   of transferred nucleon number. The horizontal axis is the impact 
   parameter $b$. Results for the reactions at $E_\mathrm{lab}=$ 
   235~MeV are denoted by triangles, while results for the reactions 
   at $E_\mathrm{lab}=$ 249~MeV are denoted by circles. In (b), 
   the equilibrium $N/Z$ value of the total system, 1.43, is indicated by 
   a horizontal dashed line.
   }
   \label{FIG:Nave(40Ca+208Pb)}
\end{figure}
%&&&&&&&&&&&&&&&&&&&&&&&&&&&&&&&&&&&&&&&&&&&&&&&&&&

Figure~\ref{FIG:Nave(40Ca+208Pb)} shows the average number of 
transferred nucleons in (a), the $N/Z$ ratios of the PLF and the TLF in (b), 
the average number of nucleons emitted to the continuum in (c), 
and the fluctuation of the transferred nucleon number in (d), as 
functions of impact parameter. In each panel, triangles represent results 
for $E_\mathrm{lab}=$ 235~MeV and circles represent results for 
$E_\mathrm{lab}=$ 249~MeV. 

In Fig.~\ref{FIG:Nave(40Ca+208Pb)}~(a), the average number of 
transferred neutrons is shown by filled symbols connected with solid 
lines, while the average number of transferred protons is shown by 
open symbols connected with dotted lines. Positive values indicate 
the increase of the projectile nucleons (transfer from $^{208}$Pb 
to $^{40}$Ca) and negative values indicate the decrease (transfer 
from $^{40}$Ca to $^{208}$Pb). As seen from the figure, 
the average number of transferred protons shows a minimum at 
a certain impact parameter ($b=$ 4.0~fm for $E_\mathrm{lab}=$ 
235~MeV and $b=$ 5.0~fm for $E_\mathrm{lab}=$ 249~MeV).
Outside this impact parameter, the nucleon transfer process 
proceeds towards the direction of the charge equilibrium of the 
projectile and the target. Inside this impact parameter, neutrons 
are still transferred towards the same direction. However, the 
number of transferred protons decreases and becomes positive,
which corresponds to the transfer from $^{208}$Pb to 
$^{40}$Ca.

At first sight, the direction of the proton transfer at small impact 
parameter region is opposite to the direction of the charge equilibrium. 
However, it is not the case as can be understood from 
Fig.~\ref{FIG:Nave(40Ca+208Pb)}~(b) which shows the 
neutron-to-proton ratios, $N/Z$, of the PLF (filled symbols 
connected with solid lines) and the TLF (open symbols 
connected with dotted lines), which are obtained from the 
average numbers of the nucleons shown in 
Fig.~\ref{FIG:Nave(40Ca+208Pb)}~(a). Before collision, 
the $N/Z$ ratio is given by 1.00 for $^{40}$Ca and 
1.54 for $^{208}$Pb. In Fig.~\ref{FIG:Nave(40Ca+208Pb)}~(b), 
the $N/Z$ ratio of the total system, 1.43, is shown by a horizontal 
dashed line. As seen from the figure, the nucleon transfer processes 
proceed towards the direction of the charge equilibrium for both 
the PLF and the TLF at all impact parameter region outside the fusion 
critical impact parameter. Even though the average number of 
transferred protons shows complex behavior at small impact 
parameter region, the $N/Z$ ratios of the PLF and the TLF 
monotonically approach to the fully equilibrated value of 1.43 
as the impact parameter decreases.

The change of sign of the average number of transferred protons 
at small impact parameter region is found to be related to the 
formation of a rather solid neck. When the neck is broken, we 
find that most part of the neck is absorbed by the lighter fragment 
({\it cf.} Fig.~\ref{FIG:TDPLOT(40Ca+208PbE249,b4.56)}). 
Since the neck is composed of both neutrons and protons, the 
absorption of the nucleons in the neck region results in the 
increase of average number of nucleons in the PLF for both 
neutrons and protons (see Fig.~\ref{FIG:Nave(40Ca+208Pb)}~(a)).

In Fig.~\ref{FIG:Nave(40Ca+208Pb)}~(c), we show the average 
number of nucleons emitted to the continuum during the time evolution. 
The average number of neutrons (protons) emitted to the continuum 
is denoted by filled (open) symbols connected with solid (dotted) lines.
We count it by subtracting the number of nucleons inside a sphere 
of 14~fm for the TLF and that inside a sphere of 10~fm for the PLF from 
the total number of nucleons, 248. The average number of emitted nucleons is 
again very small, at most 0.1 around the fusion critical impact parameter.

In Fig.~\ref{FIG:Nave(40Ca+208Pb)}~(d), we show the fluctuation
of the transferred nucleon number. The fluctuation of transferred 
neutron (proton) number is denoted by filled (open) symbols 
connected with solid (dotted) lines. The fluctuation increases 
monotonically as the impact parameter decreases, reaching the 
maximum value roughly 1.3 around the fusion critical impact parameter.
Although the average number of transferred protons is small at 
the small impact parameter region, the fluctuation of transferred 
proton number has value as large as that of neutrons. This fact 
indicates that single-particle wave functions of protons are exchanged 
actively between the projectile and the target, although the number 
of transferred protons is small on average.

\subsubsection{Transfer probabilities}

%&&&&&&&&&&&&&&&&&&&&&&&&&&&&&&&&&&&&&&&&&&&&&&&&&&
\begin{figure}[t]
   \begin{center}
   \includegraphics[width=8.6cm]{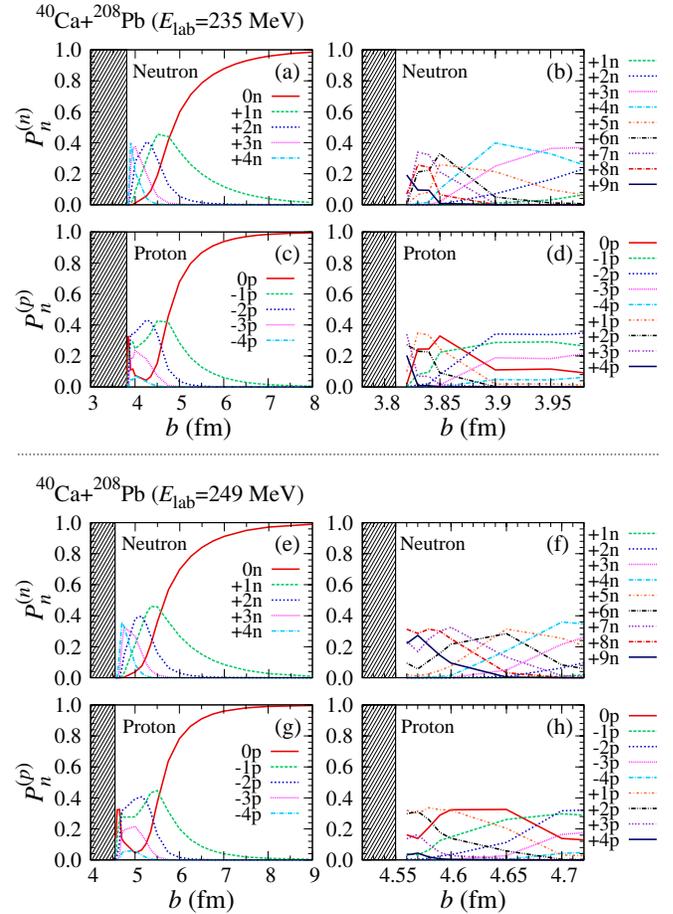}
   \end{center}\vspace{-3mm}
   \caption{(Color online)
   Neutron and proton transfer probabilities as functions of impact 
   parameter $b$ for the \RE{40}{Ca}+{208}{Pb} reactions. 
   (a), (b), (c), and (d): Results at $E_\mathrm{lab}=$ 235~MeV.
   (e), (f), (g), and (h): Results at $E_\mathrm{lab}=$ 249~MeV.
   The positive (negative) number of transferred nucleons 
   represents the number of nucleons added to (removed from) the 
   projectile. Note that horizontal scales are different between 
   the left and the right panels. Shaded regions at small 
   impact parameter ($b\le3.81$~fm for $E_\mathrm{lab}=$ 235~MeV 
   and $b\le4.55$~fm for $E_\mathrm{lab}=$ 249~MeV) correspond 
   to the fusion reactions.
   }
   \label{FIG:Pn(40Ca+208Pb)}
\end{figure}
%&&&&&&&&&&&&&&&&&&&&&&&&&&&&&&&&&&&&&&&&&&&&&&&&&&

The nucleon transfer probabilities, $P_n^{(q)}(b)$, are shown 
in Fig.~\ref{FIG:Pn(40Ca+208Pb)}. Top panels 
((a), (b), (c), and (d)) show results at $E_\mathrm{lab}=$ 235 MeV, 
while lower panels ((e), (f), (g), and (h)) show results at 
$E_\mathrm{lab}=$ 249 MeV. In the figure, shaded regions 
at small impact parameter ($b \le 3.81$~fm for $E_\mathrm{lab}=$ 
235~MeV and $b \le 4.55$~fm for $E_\mathrm{lab}=$ 249~MeV) 
correspond to the fusion reactions. The positive (negative) number of 
transferred nucleons represents the number of nucleons added to 
(removed from) the projectile. In the left panels ((a), (c), (e), and (g)), 
we show transfer probabilities, ($0$n) to ($+4$n) for neutrons and 
($0$p) to ($-4$p) for protons. In the right panels ((b), (d), (f), and (h)), 
we show transfer probabilities, ($+1$n) to ($+9$n) for neutrons and 
($-4$p) to ($+4$p) for protons, at small impact parameter regions 
just outside the fusion critical impact parameter. Probabilities of 
neutron transfer from $^{40}$Ca to $^{208}$Pb are very small 
and are not shown.

As in the case of \RE{40,\,48}{Ca}+{124}{Sn} reactions, we find 
that probabilities of single-nucleon transfer (green dashed lines) 
extend to a large impact parameter region. Reaction probabilities 
for multinucleon transfer processes become appreciable at a small 
impact parameter region close to the fusion critical impact parameter. 
The transfer probabilities towards the charge equilibrium are large in 
most cases. At a small impact parameter region just outside the fusion 
critical impact parameter, however, we find substantial probabilities 
for the proton transfer processes opposite to the charge equilibrium 
as seen in the right panels of Fig.~\ref{FIG:Pn(40Ca+208Pb)} 
((b), (d), (f), and (h)). This is related to the increase of the average 
number of transferred protons at small impact parameter region 
which was seen in Fig.~\ref{FIG:Nave(40Ca+208Pb)}~(a).

\subsubsection{Transfer cross sections}

%&&&&&&&&&&&&&&&&&&&&&&&&&&&&&&&&&&&&&&&&&&&&&&&&&&
\begin{figure}[t]
   \begin{center}
   \includegraphics[width=8.6cm]{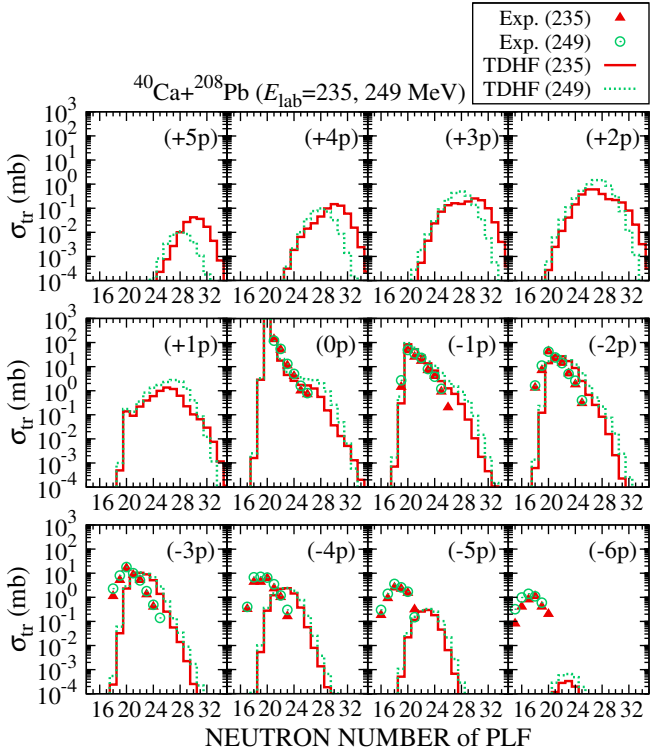}
   \end{center}\vspace{-3mm}
   \caption{(Color online)
   Transfer cross sections for the \RE{40}{Ca}+{208}{Pb} reactions 
   at $E_\mathrm{lab}=$ 235 and 249~MeV. Red filled triangles (green 
   open circles) denote measured cross sections at $E_\mathrm{lab}=$ 
   235 (249) MeV. Red solid (green dotted) lines denote results of the 
   TDHF calculations at $E_\mathrm{lab}=$ 235 (249) MeV. The number 
   of transferred protons (positive number for the transfer from  
   $^{208}$Pb to $^{40}$Ca) is indicated as ($x$p) ($-6 \le x \le +5$). 
   The measured cross sections have been reported in
   Ref.~\cite{Szilner(40Ca+208Pb)2}
   }
   \label{FIG:NTCS(40Ca+208Pb,vs-N)}
\end{figure}
%&&&&&&&&&&&&&&&&&&&&&&&&&&&&&&&&&&&&&&&&&&&&&&&&&&

%&&&&&&&&&&&&&&&&&&&&&&&&&&&&&&&&&&&&&&&&&&&&&&&&&&
\begin{figure}[t]
   \begin{center}
   \includegraphics[width=8.6cm]{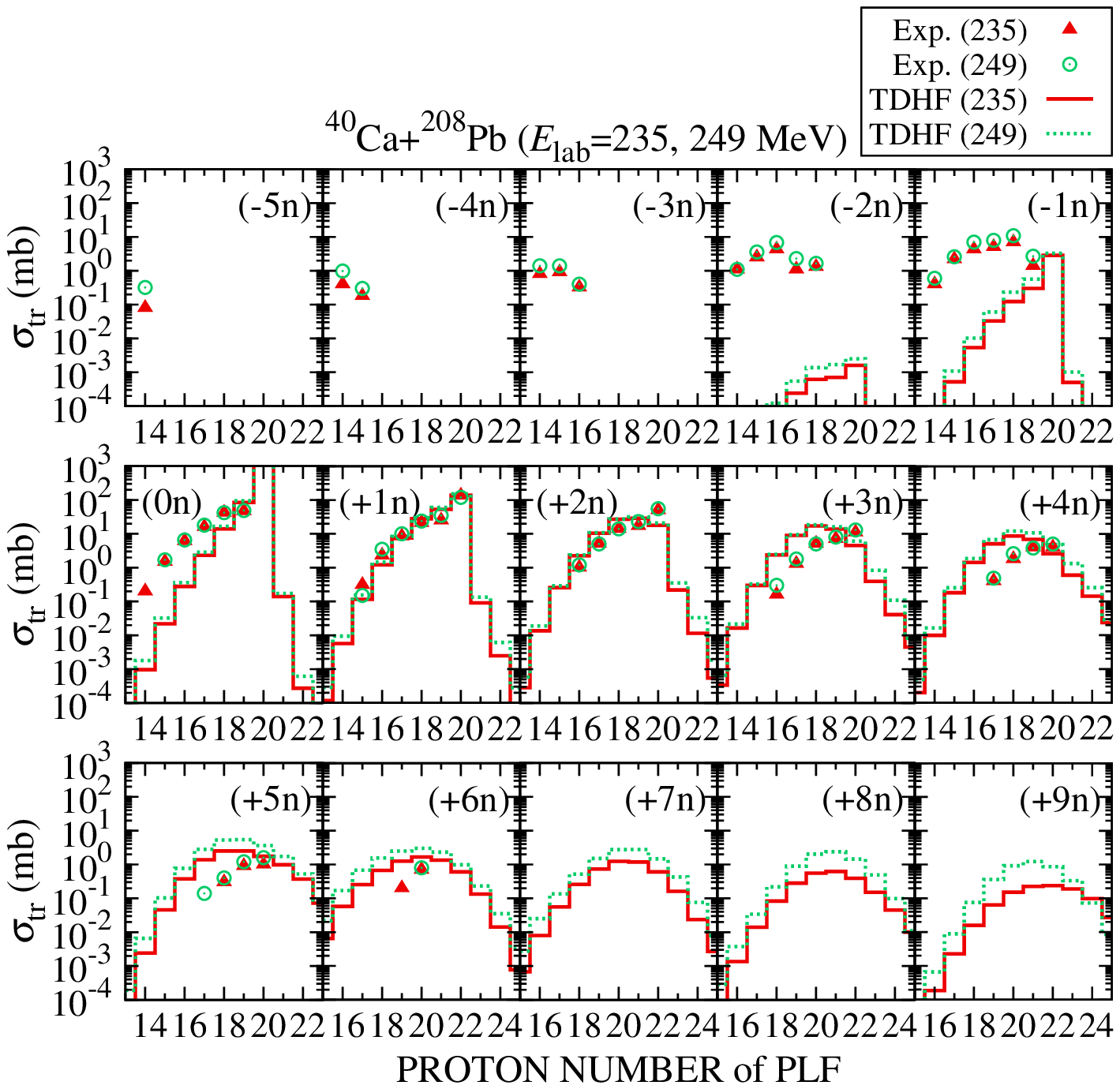}
   \end{center}\vspace{-3mm}
   \caption{(Color online)
   The same transfer cross sections for the \RE{40}{Ca}+{208}{Pb} 
   reactions as those in Fig. 14. The number of transferred neutrons is 
   indicated as ($x$n) ($-5 \le x \le +9$).
   }
   \label{FIG:NTCS(40Ca+208Pb,vs-Z)}
\end{figure}
%&&&&&&&&&&&&&&&&&&&&&&&&&&&&&&&&&&&&&&&&&&&&&&&&&&

We show transfer cross sections in Figs.~\ref{FIG:NTCS(40Ca+208Pb,vs-N)} 
and \ref{FIG:NTCS(40Ca+208Pb,vs-Z)}. Each panel of 
Fig.~\ref{FIG:NTCS(40Ca+208Pb,vs-N)} shows cross sections 
classified according to the change of the proton number of the PLF from 
$^{40}$Ca which is indicated by ($x$p) ($-6 \le x \le +5$), as functions 
of neutron number of the PLF. Each panel of 
Fig.~\ref{FIG:NTCS(40Ca+208Pb,vs-Z)} shows cross sections 
classified according to the change of the neutron number of the PLF
from $^{40}$Ca which is indicated by ($x$n) ($-5 \le x \le +9$), 
as functions of proton number of the PLF. Red filled triangles denote 
measured cross sections for $E_\mathrm{lab}=$ 235~MeV, while 
green open circles denote those for $E_\mathrm{lab}=$ 249~MeV. 
Cross sections calculated by the TDHF are denoted by red solid 
(green dotted) lines for $E_\mathrm{lab}=$ 235 (249) MeV. 
As seen in the average number of transferred nucleons in 
Fig.~\ref{FIG:Nave(40Ca+208Pb)}~(a) and in the transfer 
probabilities in Fig.~\ref{FIG:Pn(40Ca+208Pb)}, the transfer cross 
sections towards the direction of the charge equilibrium dominate.

In ($0$p) and ($-1$p) panels of 
Fig.~\ref{FIG:NTCS(40Ca+208Pb,vs-N)}, the TDHF calculation 
is seen to reproduce the measured cross sections up to six-neutron 
transfer. As the number of transferred protons increases, ($-2$p) to 
($-6$p), the cross sections in the TDHF calculation show a maximum 
at a neutron number more than that of $^{40}$Ca. Compared with 
measured cross sections, the TDHF results shift towards larger values of 
neutron number. This behavior is similar to the case of 
\RE{40}{Ca}+{124}{Sn} reaction. Looking at the transfer 
cross sections for a fixed number of transferred neutrons in 
Fig.~\ref{FIG:NTCS(40Ca+208Pb,vs-Z)}, the TDHF calculations 
reproduce ($+1$n) and ($+2$n) panels rather well. 

As seen in Fig.~\ref{FIG:NTCS(40Ca+208Pb,vs-N)}, the TDHF 
calculations provide substantial cross sections for proton pickup 
reactions, ($+1$p) to ($+5$p), which is the transfer towards the 
opposite direction of the charge equilibrium expected from the 
initial $N/Z$ ratios. The cross sections show a peak around the 
neutron number 28. The TDHF calculations also provide substantial 
cross sections for many neutron pickup reactions
(see bottom row of Fig.~\ref{FIG:NTCS(40Ca+208Pb,vs-Z)}).
The cross sections show a peak around the proton number 20. 
These cross sections come from an impact parameter region close to 
the fusion critical impact parameter. As seen in 
Fig.~\ref{FIG:Nave(40Ca+208Pb)}~(a), a large average 
number of transferred neutrons up to 10 is seen while the 
average number of transferred protons has small value. 
We note that the collision close to the fusion critical impact 
parameter accompanies large TKEL, and should suffer substantial 
evaporation effects which are not treated in the present analyses.

The TDHF calculation systematically underestimates the cross 
section of neutron transfer processes from $^{40}$Ca to 
$^{208}$Pb, ($-1$n) to ($-5$n) (see top row of 
Fig.~\ref{FIG:NTCS(40Ca+208Pb,vs-Z)}). Although these 
processes are against the charge equilibrium, substantial cross 
sections are observed experimentally. In the TDHF calculation, 
cross sections of neutron transfer channels opposite to the charge 
equilibrium are several orders of magnitude smaller than the 
measurements. In Ref.~\cite{Szilner(40Ca+208Pb)2}, it has been 
argued that the neutron evaporation after collision is responsible 
for these channels.

\subsection{\RE{58}{Ni}+{208}{Pb} reaction}{\label{Subsec(58Ni+208Pb)}}

As a final case, we present results for the \RE{58}{Ni}+{208}{Pb} 
reaction at $E_\mathrm{lab}=$ 328.4~MeV 
($E_\mathrm{c.m.}\simeq$ 256.8~MeV), for which measurements 
are reported in Ref.~\cite{Corradi(58Ni+208Pb)}. 
Since this system has $Z_\mathrm{P}Z_\mathrm{T}$ = 2296 
exceeding the critical value 1600, we may expect an appearance of 
the quasi-fission process at a small impact parameter region. 
Using the frozen-density approximation, the Coulomb barrier 
height is estimated to be $V_\mathrm{B}\approx$ 247.6~MeV, 
which is lower than the center-of-mass energy. We find the fusion 
critical impact parameter $b_\mathrm{f}$ given by 1.38~fm 
for this reaction. To decide whether the nucleus once gets fused 
eventually decays into fragments or not, we continue to calculate 
the time evolution up to 4000~fm/c after two nuclei touches. 
If the fused system keeps compact form for this period, we regard 
the process as fusion.

\subsubsection{Overview of the reaction}{\label{Overview(58Ni+208Pb)}}

%&&&&&&&&&&&&&&&&&&&&&&&&&&&&&&&&&&&&&&&&&&&&&&&&&&
\begin{figure}[t]
   \begin{center}
   \includegraphics[width=6.5cm]{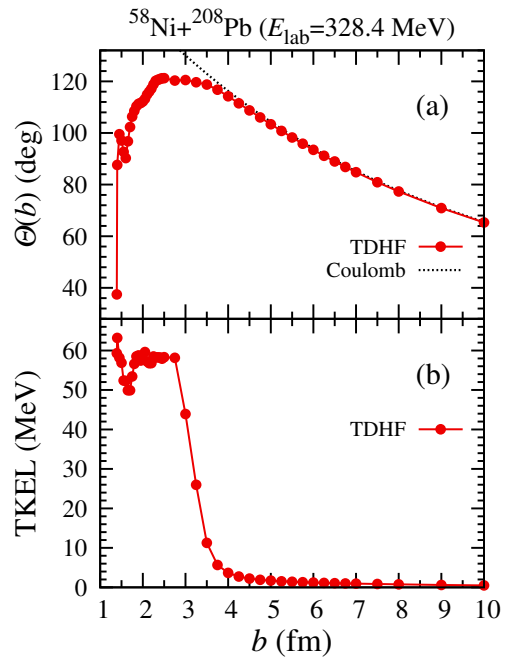}
   \end{center}\vspace{-3mm}
   \caption{(Color online)
   Deflection function (a) and total kinetic energy loss (b) as functions 
   of impact parameter $b$ for the reactions of \RE{58}{Ni}+{208}{Pb} 
   at $E_\mathrm{lab}=$ 328.4~MeV. In (a), we show a deflection 
   function for the pure Coulomb trajectory by a dotted line.
   }
   \label{FIG:Theta+TKEL(58Ni+208Pb)}
\end{figure}
%&&&&&&&&&&&&&&&&&&&&&&&&&&&&&&&&&&&&&&&&&&&&&&&&&&

%&&&&&&&&&&&&&&&&&&&&&&&&&&&&&&&&&&&&&&&&&&&&&&&&&&
\begin{figure}[b]
   \begin{center}
   \includegraphics[width=8cm]{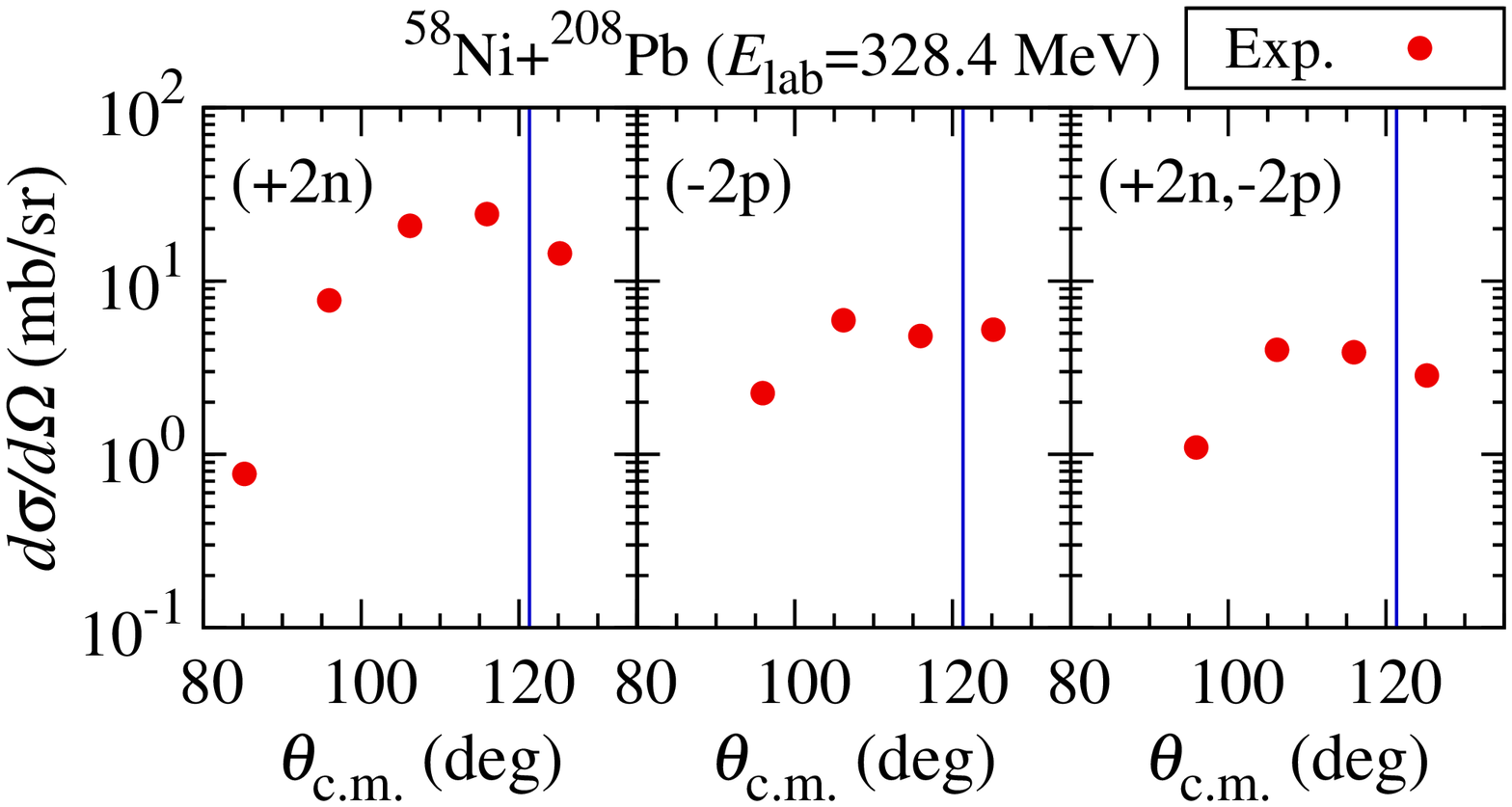}
   \end{center}\vspace{-3mm}
   \caption{(Color online)
   Differential cross sections of representative transfer 
   channels as functions of scattering angle in the center-of-mass 
   frame for the \RE{58}{Ni}+{208}{Pb} reaction at 
   $E_\mathrm{lab}=$ 328.4~MeV. The Coulomb rainbow angle 
   obtained from the TDHF trajectories is denoted by blue solid vertical 
   lines, and is compared with measured differential cross sections, 
   red filled circles, which have been reported in 
   Ref.~\cite{Corradi(58Ni+208Pb)}. 
   }
   \label{FIG:DCS(58Ni+208Pb)_exp}
\end{figure}
%&&&&&&&&&&&&&&&&&&&&&&&&&&&&&&&&&&&&&&&&&&&&&&&&&&

We first present an overview of the reaction dynamics.
In Fig.~\ref{FIG:Theta+TKEL(58Ni+208Pb)}, we show the 
deflection function in (a) and the TKEL in (b), as functions of 
impact parameter. In (a), we also show a deflection function for 
the pure Coulomb trajectory by a dotted line. The Coulomb 
rainbow occurs at the impact parameter of 2.5~fm and the 
rainbow angle is  $\theta_\mathrm{r} \simeq 121^\circ$. 
In Fig.~\ref{FIG:DCS(58Ni+208Pb)_exp}, 
we compare the Coulomb rainbow angle with measured 
differential cross sections which have been reported in 
Ref.~\cite{Corradi(58Ni+208Pb)}. Red filled circles denote 
measured cross sections and blue solid vertical lines denote 
the Coulomb rainbow angle. From the figure, we find the 
measured differential cross sections show rather flat distributions 
compared with lighter systems. This may be related to 
a rather small curvature of the deflection function around the 
Coulomb rainbow angle obtained from the TDHF trajectories as 
seen in Fig.~\ref{FIG:Theta+TKEL(58Ni+208Pb)}~(a).

The TKEL shows a behavior different from lighter systems. 
The maximum TKEL is about 50-60~MeV, similar to the value observed 
in \RE{40}{Ca}+{208}{Pb} reaction. However, there is a large 
impact parameter region, from 1.39~fm to 2.75~fm, in which 
the TKEL takes approximately the same value. 

%&&&&&&&&&&&&&&&&&&&&&&&&&&&&&&&&&&&&&&&&&&&&&&&&&&
\begin{figure}[t]
   \begin{center}
   \includegraphics[width=8.6cm]{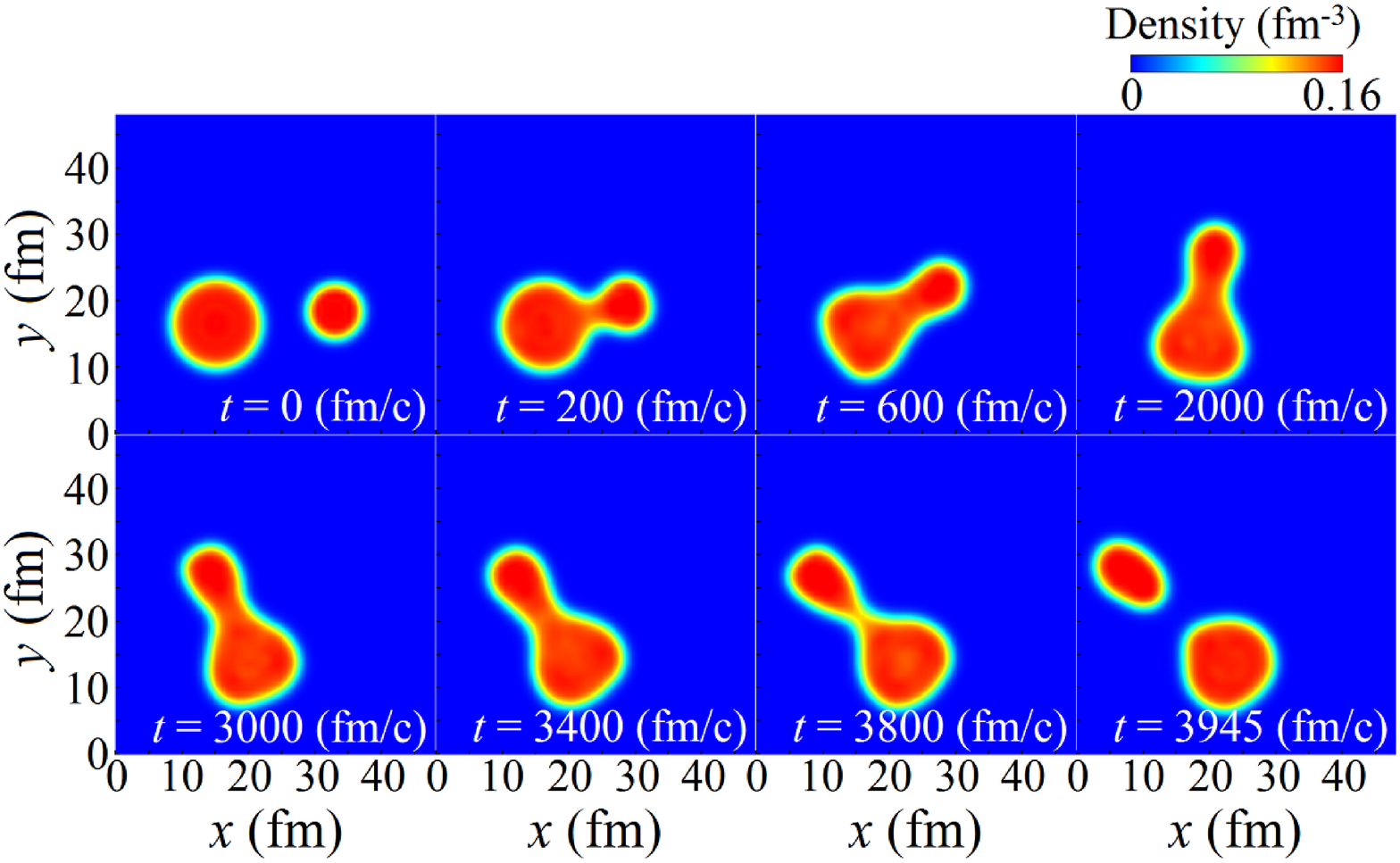}
   \end{center}\vspace{-3mm}
   \caption{(Color online)
   Snapshots of density distribution of the \RE{58}{Ni}+{208}{Pb} 
   reaction at $E_\mathrm{lab}=$ 328.4~MeV and $b=$ 1.39~fm, 
   just outside the fusion critical impact parameter.
   }
   \label{FIG:TDPLOT(58Ni+208Pb,b1.39)}
\end{figure}
%&&&&&&&&&&&&&&&&&&&&&&&&&&&&&&&&&&&&&&&&&&&&&&&&&&

In Fig.~\ref{FIG:TDPLOT(58Ni+208Pb,b1.39)}, we show snapshots
of density distribution for the \RE{58}{Ni}+{208}{Pb} reaction at 
the impact parameter of 1.39~fm, just outside the fusion critical 
impact parameter. In the course of the collision, colliding nuclei 
form a rather thick neck and excurse for a long period connected 
by the neck. We find the two nuclei are connected for a period 
as long as 3600~fm/c. For collisions in the impact parameter region 
where the TKEL takes values around 50-60~MeV, we find a formation 
of a similar thick neck which persists rather long period. These reactions 
are considered to correspond to the quasi-fission. 

%&&&&&&&&&&&&&&&&&&&&&&&&&&&&&&&&&&&&&&&&&&&&&&&&&&
\begin{figure}[t]
   \begin{center}
   \includegraphics[height=16cm]{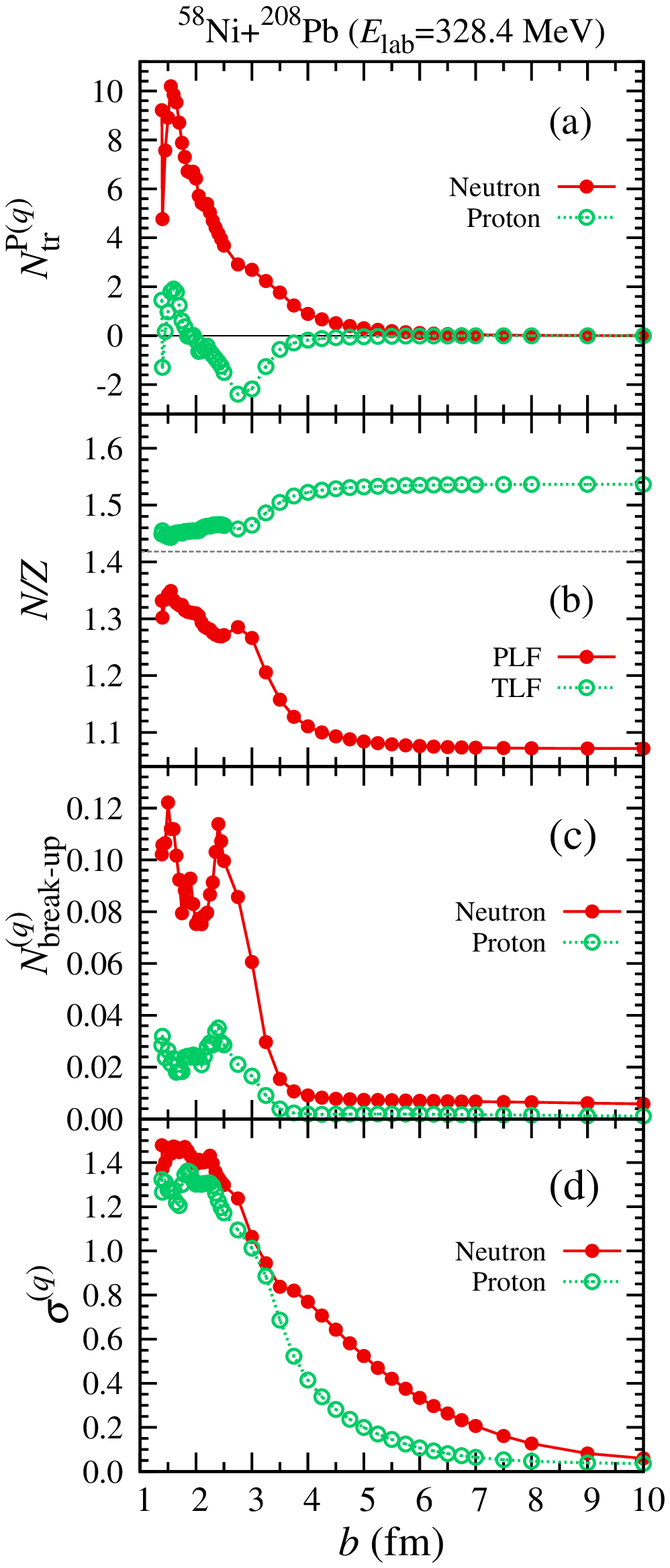}
   \end{center}\vspace{-3mm}
   \caption{(Color online)
   The \RE{58}{Ni}+{208}{Pb} reaction at $E_\mathrm{lab}=$ 
   328.4~MeV. (a): Average number of transferred nucleons from 
   the target to the projectile. (b): Neutron-to-proton ratios, $N/Z$, 
   of the PLF and the TLF after collision. (c): Average number of 
   nucleons emitted to the continuum. (d): Fluctuation of transferred 
   nucleon number. The horizontal axis is the impact parameter $b$. 
   In (b), the equilibrium $N/Z$ value of the total system, 1.42, is 
   indicated by a horizontal dashed line.
   }
   \label{FIG:Nave(58Ni+208Pb)}
\end{figure}
%&&&&&&&&&&&&&&&&&&&&&&&&&&&&&&&&&&&&&&&&&&&&&&&&&&

Figure~\ref{FIG:Nave(58Ni+208Pb)} shows the average number of 
transferred nucleons in (a), the $N/Z$ ratios of the PLF and the TLF in (b), 
the average number of nucleons emitted to the continuum in (c), and 
the fluctuation of the transferred nucleons in (d), as functions of 
impact parameter.

In Fig.~\ref{FIG:Nave(58Ni+208Pb)}~(a), the average number of
transferred neutrons is shown by red filled circles connected 
with solid lines, while the average number of transferred protons
is shown by green open circles connected with dotted lines. 
Positive numbers indicate the increase of the projectile nucleons 
(transfer from $^{208}$Pb to $^{58}$Ni) and negative
numbers indicate the decrease (transfer from $^{58}$Ni to 
$^{208}$Pb). From the figure, we find the average number of transferred
protons shows a minimum at $b=$ 2.75~fm. We note that this value
coincides with the impact parameter inside which the TKEL becomes
almost constant in Fig.~\ref{FIG:Theta+TKEL(58Ni+208Pb)}~(b). 
A similar minimum was also seen in the \RE{40}{Ca}+{208}{Pb} 
case, as shown in Fig.~\ref{FIG:Nave(40Ca+208Pb)}~(a). 
Outside this impact parameter, nucleons are transferred towards 
the direction of the charge equilibrium expected from the initial 
$N/Z$ ratios. In the impact parameter region, 1.55~fm $\le b \le$ 
2.75~fm, the average number of transferred nucleons increases as 
the impact parameter decreases for both neutrons and protons. 
A similar behavior was also seen in \RE{40}{Ca}+{208}{Pb} 
reaction as in Fig.~\ref{FIG:Nave(40Ca+208Pb)}~(a). 

At the impact parameter region $b<$ 1.85~fm, the average number 
of transferred protons becomes positive, opposite to the direction of 
the charge equilibrium of the initial system. However, the nucleon 
transfer still proceeds towards the charge equilibrium of both the 
PLF and the TLF after the collision. This is clearly seen 
in Fig.~\ref{FIG:Nave(58Ni+208Pb)}~(b) which shows the $N/Z$ 
ratios of the PLF and the TLF after collision. The $N/Z$ ratio of the PLF 
(TLF) is denoted by red filled (green open) circles connected with 
solid (dotted) lines. As seen from the figure, the $N/Z$ ratios of 
both the PLF and the TLF become closer to the $N/Z$ ratio of the 
total system, 1.42, which is represented by a horizontal dashed line. 

As mentioned in the case of \RE{40}{Ca}+{208}{Pb} reactions, 
the change in the average number of transferred protons across 
the impact parameter $b \sim 3$~fm is related to the formation 
of the neck. Outside $b \sim 3$~fm, the neck is not formed and 
two nuclei are separated even at the closest approach. In such case, 
nucleons are transferred towards the direction of the charge 
equilibrium expected from the initial $N/Z$ ratios.
Inside $b \sim 3$~fm, the neck is formed between two nuclei. 
Then the transfer of nucleons proceeds in two steps. Before the 
formation of the neck, the transfer of nucleons proceeds 
towards the charge equilibrium of the initial system in the 
same way as that in $b>3$~fm. After the formation of 
the neck, an exchange of a large number of nucleons occurs at 
the time of the breaking of the neck. Depending on the position 
of the neck breaking, the transfer of nucleons is expected in 
either directions, from the target to the projectile or the reverse. 
Since the neck is formed with both protons and neutrons, the 
nucleon transfer in the neck breaking process accompanies both 
protons and neutrons in the same direction.

Looking at Fig.~\ref{FIG:Nave(58Ni+208Pb)}~(a), we find 
the increase of the average numbers of transferred nucleons of 
both neutrons and protons as the impact parameter decreases 
below $b=2.75$~fm. This indicates that the neck is broken at 
the position close to the target. Both protons and neutrons in the 
neck region are absorbed by the projectile. This mechanism explains 
the reason why the number of transferred protons increases as the 
impact parameter decreases in Fig.~\ref{FIG:Nave(58Ni+208Pb)}~(a). 
This transfer process associated with the neck breaking was also 
seen in the \RE{40,\,48}{Ca}+{124}{Sn} and the 
\RE{40}{Ca}+{208}{Pb} reactions.

At very small impact parameter region, 1.40~fm $\le b \le$ 
1.50~fm, the average number of transferred neutrons shows 
a large fluctuation. The average number of transferred protons 
also shows the fluctuation, correlated with that of neutrons. 
These fluctuations occur by changes of the breaking point of the neck. 
When the neck is broken close to the target, a large number of 
nucleons are transferred from the target to the projectile, while 
the neck is broken at a midpoint between the projectile and the 
target, the number of transferred nucleons becomes small.

Figure~\ref{FIG:Nave(58Ni+208Pb)}~(c) shows the average
number of nucleons emitted to the continuum during 
the time evolution. The average number of neutrons (protons)
emitted to the continuum is denoted by red filled (green open) 
circles connected with solid (dotted) lines. As in other systems, 
we calculate it by subtracting the average number of nucleons 
inside a sphere of 14~fm for the TLF and that inside a sphere 
of 10~fm for the PLF from the total number of nucleons, 266.
Again the number is rather small, about 0.12 at the maximum. 

Figure~\ref{FIG:Nave(58Ni+208Pb)}~(d) shows the fluctuation 
of the transferred nucleon number. The fluctuation of transferred 
neutron (proton) number is denoted by red filled (green open) 
circles connected with solid (dotted) lines. They show a different 
behavior across the impact parameter around 3~fm, indicating 
a qualitative change of the dynamics. Outside this impact 
parameter where protons and neutrons are transferred 
in different directions, the fluctuation of transferred neutron number 
is larger than that of protons. Inside this impact parameter, although 
the average number of transferred neutrons is much larger than 
that of protons, the fluctuation is almost the same. This 
indicates that although the average number of transferred 
protons is small, there is a strong mixture of single-particle orbitals 
of protons due to the formation and breaking of the neck.

\subsubsection{Transfer probabilities}

%&&&&&&&&&&&&&&&&&&&&&&&&&&&&&&&&&&&&&&&&&&&&&&&&&&
\begin{figure}[b]
   \begin{center}
   \includegraphics[width=8.6cm]{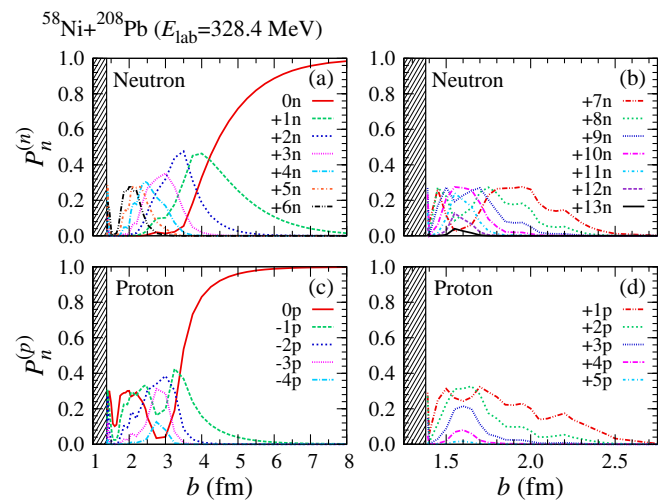}
   \end{center}\vspace{-3mm}
   \caption{(Color online)
   Neutron and proton transfer probabilities as functions of impact 
   parameter $b$ for the reactions of \RE{58}{Ni}+{208}{Pb} at 
   $E_\mathrm{lab}=$ 328.4~MeV. Figure (a) and (b) show probabilities
   of neutrons, while figure (c) and (d) show those of protons. The positive 
   (negative) number of transferred nucleons represents the number of 
   nucleons added to (removed from) the projectile. Note that horizontal 
   scales are different between the left and the right panels. A shaded 
   region at small impact parameter ($b\le1.38$~fm) corresponds 
   to the fusion reactions.
   }
   \label{FIG:Pn(58Ni+208Pb)}
\end{figure}
%&&&&&&&&&&&&&&&&&&&&&&&&&&&&&&&&&&&&&&&&&&&&&&&&&&

We next show transfer probabilities of the \RE{58}{Ni}+{208}{Pb} 
reaction as functions of impact parameter, which are shown in 
Fig.~\ref{FIG:Pn(58Ni+208Pb)}. The small impact parameter region 
($b \le 1.38$~fm) corresponding to the fusion reaction are shaded. 
The positive (negative) number of the transferred nucleons represents 
the number of nucleons added to (removed from) the projectile.
The upper (lower) panels show neutron (proton) transfer probabilities 
for each transfer channel. In the left panels ((a) and (c)), we show transfer 
probabilities, ($0$n) to ($+6$n) for neutrons and ($0$p) to ($-4$p) 
for protons. They correspond to the transfer processes towards the 
charge equilibrium of the initial system. In the right panels ((b) and (d)), 
we show transfer probabilities, ($+7$n) to ($+13$n) for neutrons and 
($+1$p) to ($+5$p) for protons, which dominate in the small impact 
parameter region, $b\le$ 2.75~fm. Probabilities of neutron transfer 
from $^{58}$Ni to $^{208}$Pb are very small and are not shown.

In contrast to the previous cases of \RE{40,\,48}{Ca}+{124}{Sn} and 
\RE{40}{Ca}+{208}{Pb}, probabilities of transfer processes involving 
more than 6 neutrons are seen in rather wide impact parameter region, 
1.39~fm $\le b \le$ 2.75~fm, where the formation of the thick neck is 
observed. In Fig.~\ref{FIG:Theta+TKEL(58Ni+208Pb)}~(b), 
a large value of TKEL was also seen in the impact parameter region 
of $b<3$~fm, indicating the significance of the evaporation effects.

%&&&&&&&&&&&&&&&&&&&&&&&&&&&&&&&&&&&&&&&&&&&&&&&&&&
\begin{figure}[t]
   \begin{center}
   \includegraphics[width=8.6cm]{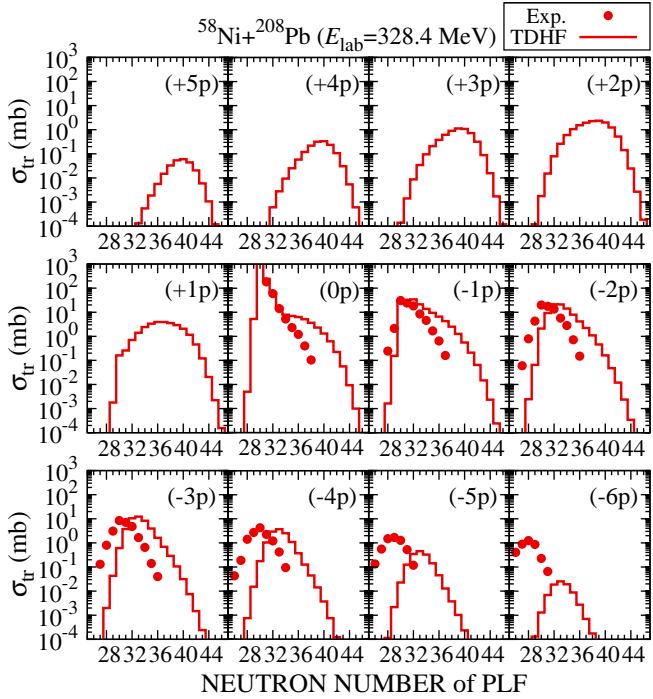}
   \end{center}\vspace{-3mm}
   \caption{(Color online)
   Transfer cross sections for the \RE{58}{Ni}+{208}{Pb} reaction 
   at $E_\mathrm{lab}=$ 328.4~MeV. Red filled circles denote 
   measured cross sections and red solid lines denote results of the TDHF 
   calculations. The number of transferred protons (positive number 
   for the transfer from $^{208}$Pb to $^{58}$Ni) is indicated as 
   ($x$p) ($-6 \le x \le +5$). The measured cross sections have been 
   reported in Ref.~\cite{Corradi(58Ni+208Pb)}.
   }
   \label{FIG:NTCS(58Ni+208Pb,vs-N)}
\end{figure}
%&&&&&&&&&&&&&&&&&&&&&&&&&&&&&&&&&&&&&&&&&&&&&&&&&&

As in previous cases, probabilities of the processes accompanying
small number of exchanged nucleons show large spatial tail.
The transfer probabilities for channels towards the charge equilibrium 
are large in most cases. The zero-proton transfer probability ($0$p, 
red solid line) in Fig.~\ref{FIG:Pn(58Ni+208Pb)}~(c) decreases as 
the impact parameter decreases, shows minimum at $b \sim 3$~fm, 
and again increases at smaller impact parameter region. This behavior 
is consistent with the behavior of the average number of transferred 
protons seen in Fig.~\ref{FIG:Nave(58Ni+208Pb)}~(a). Although 
neutron transfer probabilities to the direction opposite to the charge 
equilibrium of the initial system are vanishingly small, we find 
appreciable probabilities of proton transfer opposite to the charge 
equilibrium of the initial system, as is seen from 
Fig.~\ref{FIG:Pn(58Ni+208Pb)}~(d). This feature is again consistent 
with the behavior of the average number of transferred protons 
shown in Fig.~\ref{FIG:Nave(58Ni+208Pb)}~(a).

\subsubsection{Transfer cross sections}

%&&&&&&&&&&&&&&&&&&&&&&&&&&&&&&&&&&&&&&&&&&&&&&&&&&
\begin{figure}[t]
   \begin{center}
   \includegraphics[width=8.6cm]{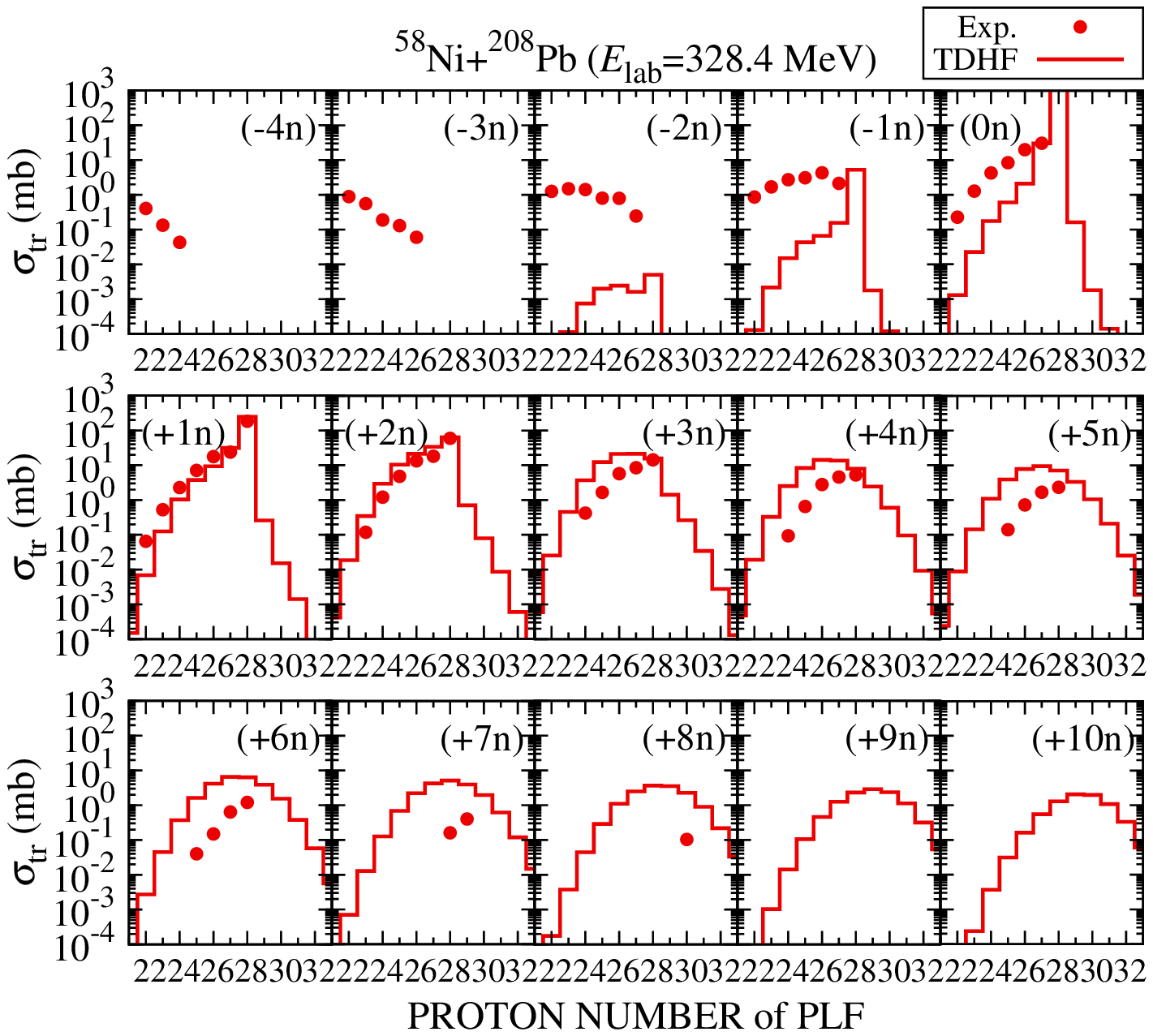}
   \end{center}\vspace{-3mm}
   \caption{(Color online)
   The same transfer cross sections for the \RE{58}{Ni}+{208}{Pb} 
   reaction as those in Fig. 21. The number of transferred neutrons is 
   indicated as ($x$n) ($-4 \le x \le +10$).
   }
   \label{FIG:NTCS(58Ni+208Pb,vs-Z)}
\end{figure}
%&&&&&&&&&&&&&&&&&&&&&&&&&&&&&&&&&&&&&&&&&&&&&&&&&&

We show transfer cross sections in Fig.~\ref{FIG:NTCS(58Ni+208Pb,vs-N)}
and Fig.~\ref{FIG:NTCS(58Ni+208Pb,vs-Z)}. Each panel of 
Fig.~\ref{FIG:NTCS(58Ni+208Pb,vs-N)} shows cross sections classified 
according to the change of the proton number of the PLF from $^{58}$Ni, 
as functions of neutron number of the PLF. Each panels of 
Fig.~\ref{FIG:NTCS(58Ni+208Pb,vs-Z)} shows cross sections classified 
according to the change of the neutron number of the PLF from $^{58}$Ni, 
as functions of proton number of the PLF. Red filled circles denote measured 
cross sections and red solid lines denote results of the TDHF calculations.
Again, reaction cross sections with relatively large values, such as 
($0$p) and ($-1$p) panels of Fig.~\ref{FIG:NTCS(58Ni+208Pb,vs-N)} 
and ($+1$n) and ($+2$n) panels of Fig.~\ref{FIG:NTCS(58Ni+208Pb,vs-Z)}, 
are described reasonably well by the TDHF calculation.

In Fig.~\ref{FIG:NTCS(58Ni+208Pb,vs-N)}, as the transferred 
proton number increases, the calculation underestimates the 
measured cross section. The peak position of the cross section 
shifts towards larger neutron number compared with the 
measurements. A similar behavior was also seen in other systems. 
This discrepancy is considered to be partly originated from neutron 
evaporation processes which we have not yet taken into account. 

In ($0$p) and ($-1$p) panels of Fig.~\ref{FIG:NTCS(58Ni+208Pb,vs-N)}, 
the TDHF calculations overestimate the cross section for channels 
accompanying large number of transferred neutrons (neutron 
number of PLF more than 34). We also find abundant cross 
sections for ($+1$p) to ($+5$p) processes, opposite to the 
charge equilibrium direction for the initial system and 
accompanying a large number of transferred neutrons. They 
come from reactions at small impact parameter region, 
$b<3$~fm, in which the transfer of nucleons associated 
with the neck breaking is appreciable.

In Fig.~\ref{FIG:NTCS(58Ni+208Pb,vs-Z)}, we find an 
underestimation of cross sections for negative neutron transfer 
($-x$n) (neutron transfer from $^{58}$Ni to $^{208}$Pb).
On the other hand, almost constant cross sections are obtained
for positive neutron transfer ($+x$n) (from $^{208}$Pb to 
$^{58}$Ni), up to the transfer of 10 neutrons. The underestimation 
of the negative neutron transfer channels may be explained by the 
evaporation effects as discussed in Ref.~\cite{Corradi(58Ni+208Pb)}. 
The cross sections for the positive neutron transfer channels originate 
from reactions at small impact parameter which accompany large 
TKEL. Therefore, they may also suffer the evaporation effects.

\section{COMPARISON WITH OTHER CALCULATIONS}{\label{Sec:Comparison}}

In this section, we compare our results of the TDHF calculations 
with those by other theories. Multinucleon transfer cross sections 
have been extensively and successfully analyzed by direct reaction 
theories such as GRAZING \cite{GRAZING} and CWKB \cite{CWKB}.
In both theories, relative motion of colliding nuclei is treated
in the semiclassical approximation. The probabilities of the 
multinucleon transfer processes are treated with a statistical 
assumption using single-particle transfer probabilities evaluated 
with the time-dependent perturbation theory. We compare our 
results with those of the GRAZING for \RE{40,\,48}{Ca}+{124}{Sn} 
reactions which have been reported in 
Refs.~\cite{Corradi(40Ca+124Sn),Corradi(48Ca+124Sn)}.

\begin{figure}[t]
   \begin{center}
   \includegraphics[width=8.6cm]{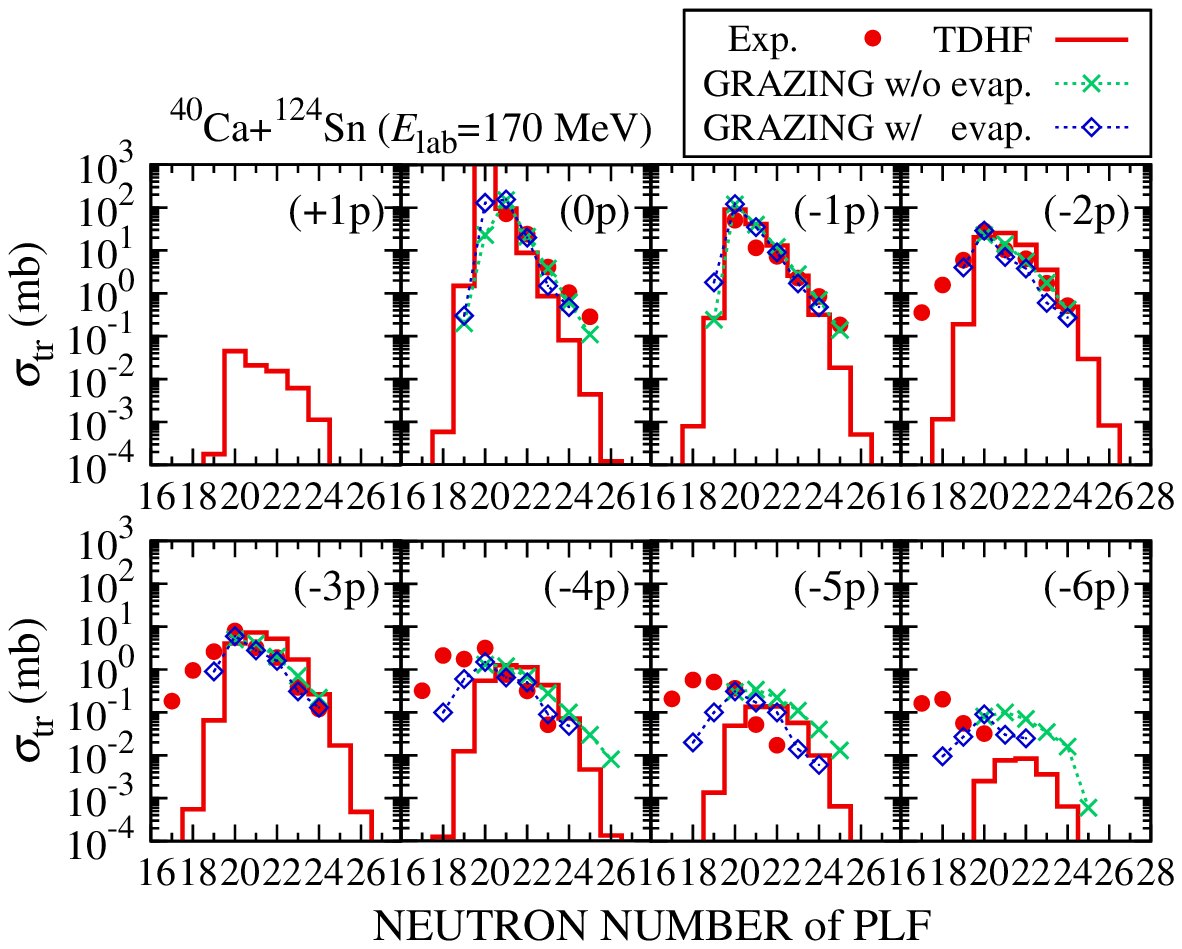}
   \end{center}\vspace{-3mm}
   \caption{(Color online)
   Transfer cross sections for the \RE{40}{Ca}+{124}{Sn} reaction 
   at $E_\mathrm{lab}=$170~MeV. Red filled circles denote 
   measured cross sections, red solid lines denote results of the 
   TDHF calculations, and green crosses (blue open diamonds) 
   connected with dotted lines denote calculated results using the 
   GRAZING code without (with) the neutron evaporation effect. 
   The number of transferred protons is indicated as ($x$p) 
   ($-6 \le x \le +1$). The measured cross sections and the GRAZING 
   results have been reported in Ref.~\cite{Corradi(40Ca+124Sn)}.
   }
   \label{FIG:NTCS(40Ca+124Sn)G}
\end{figure}

\begin{figure}[b]
   \begin{center}
   \includegraphics[width=8.6cm]{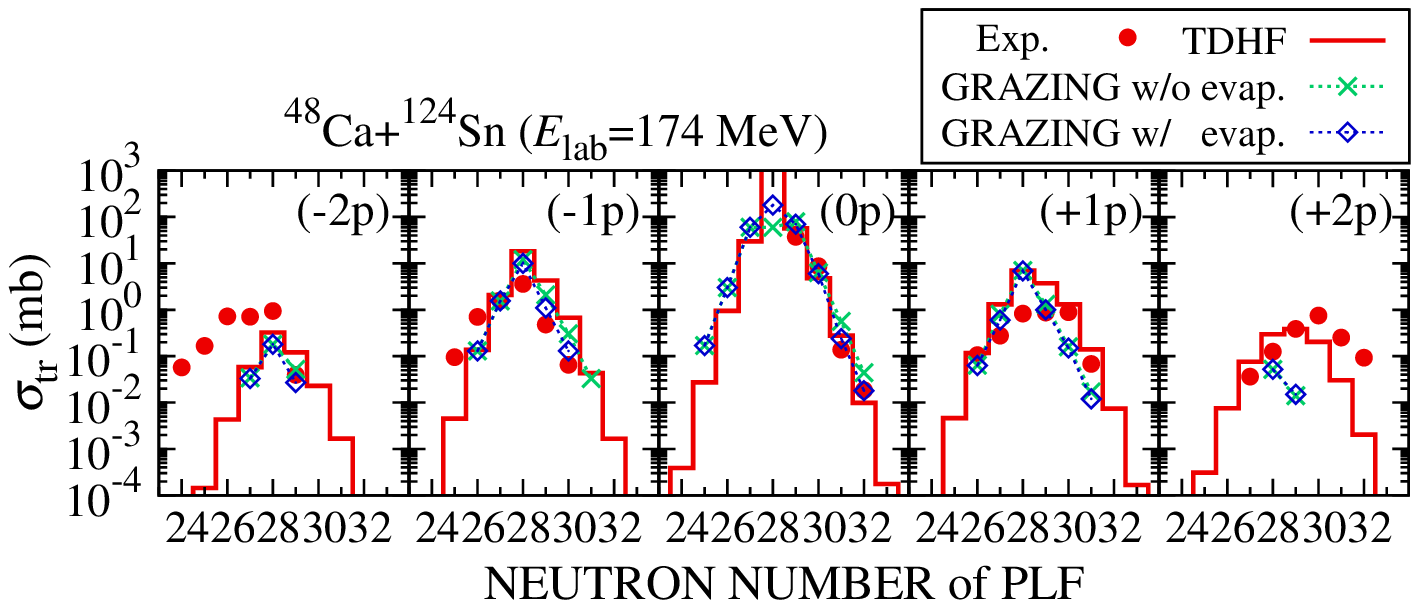}
   \end{center}\vspace{-3mm}
   \caption{(Color online)
   Transfer cross sections for the \RE{48}{Ca}+{124}{Sn} reaction 
   at $E_\mathrm{lab}=$174~MeV. Red filled circles denote 
   measured cross sections, red solid lines denote results of the 
   TDHF calculations, and green crosses (blue open diamonds) 
   connected with dotted lines denote calculated results using the 
   GRAZING code without (with) the neutron evaporation effect. 
   The number of transferred protons is indicated as ($x$p) 
   ($-2 \le x \le +2$). The measured cross sections and the GRAZING 
   results have been reported in Ref.~\cite{Corradi(48Ca+124Sn)}.
   }
   \label{FIG:NTCS(48Ca+124Sn)G}
\end{figure}

In Figs.~\ref{FIG:NTCS(40Ca+124Sn)G} and \ref{FIG:NTCS(48Ca+124Sn)G}, 
we show transfer cross sections for the reactions of 
\RE{40}{Ca}+{124}{Sn} and \RE{48}{Ca}+{124}{Sn}, respectively. 
Each panel of these figures show cross sections for transfer channels 
classified according to the change of the proton number of the PLF as 
functions of neutron number of the PLF. Red filled circles denote 
measured cross sections and red solid lines denote results of our TDHF 
calculations. Green crosses and blue open diamonds connected with 
dotted lines denote results of the GRAZING calculation. The latter 
symbols, blue open diamonds, show cross sections including the neutron 
evaporation effect, while the former symbols, green crosses, 
without the evaporation effect.

%&&&&&&&&&&&&&&&&&&&&&&&&&&&&&&&&&&&&&&&&&&&&&&&&&&
%&&&&&&&&&&&&&&&&&&&&&&&&&&&&&&&&&&&&&&&&&&&&&&&&&&

For the \RE{40}{Ca}+{124}{Sn} reaction shown in 
Fig.~\ref{FIG:NTCS(40Ca+124Sn)G}, one find that the cross sections 
by our calculation and those by the GRAZING are very close to each 
other for the processes shown in the panels of ($0$p), ($-1$p), ($-2$p), 
($-3$p), ($-4$p), and ($-5$p). Cross sections accompanying many proton 
transfer, ($-6$p), are better described by the GRAZING 
compared with the TDHF. In both TDHF and GRAZING calculations, 
the peak positions of the cross sections are shifted towards large 
number of neutrons in the ($-5$p) and ($-6$p) panels. The 
discrepancy is slightly remedied by including the neutron 
evaporation effect in the GRAZING calculation.

%&&&&&&&&&&&&&&&&&&&&&&&&&&&&&&&&&&&&&&&&&&&&&&&&&&
%&&&&&&&&&&&&&&&&&&&&&&&&&&&&&&&&&&&&&&&&&&&&&&&&&&

For the \RE{48}{Ca}+{124}{Sn} reaction shown in 
Fig.~\ref{FIG:NTCS(48Ca+124Sn)G}, we again find a good
coincidence between the TDHF results and those of the GRAZING
for ($0$p), ($\pm1$p), and ($-2$p) channels. For ($+2$p) channels, 
TDHF calculation gives better description than the GRAZING.
In both TDHF and GRAZING calculations, the cross sections
shift towards the direction of small neutron number for ($+2$p)
panel compared with measurements, while towards the direction of 
large neutron number for ($-2$p) panel. 
In Ref.~\cite{Corradi(48Ca+124Sn)}, the effect of the neutron 
evaporation has been evaluated to be small for this system.

We notice that there are similar failures in the TDHF and the GRAZING
calculations for the cross sections of channels accompanying transfer 
of large number of protons. It seems that they are caused by 
a common problem, although two theories are relied upon very 
different basis. One possible origin of the failure is an insufficient 
inclusion of the correlation effects beyond the mean-field theory. 
In the TDHF calculation, the many-body wave function is always 
assumed to be a single Slater determinant and correlations beyond 
the mean-field is not included. In the GRAZING calculation, multinucleon 
transfer probabilities are evaluated from single-nucleon transfer 
probabilities with a statistical assumption, ignoring correlation 
effects among nucleons.

%&&&&&&&&&&&&&&&&&&&&&&&&&&&&&&&&&&&&&&&&&&&&&&&&&&
\begin{figure}[b]
   \begin{center}
   \includegraphics[width=8.6cm]{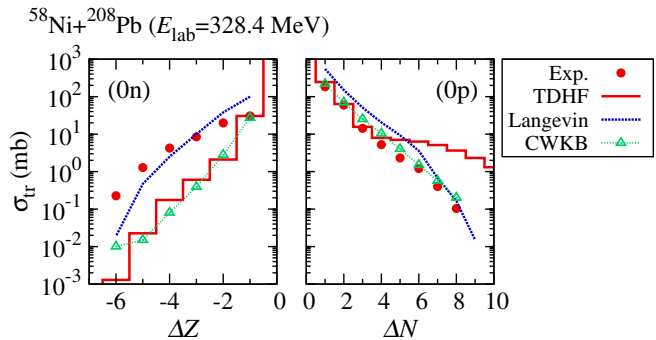}
   \end{center}\vspace{-3mm}
   \caption{(Color online)
   Cross sections for transfer channels of pure proton stripping 
   without neutron transfer (left) and pure neutron pickup without 
   proton transfer (right) for the \RE{58}{Ni}+{208}{Pb} reaction
   at $E_\mathrm{lab}=$ 328.4~MeV. Red filled circles denote 
   measured cross sections \cite{Corradi(58Ni+208Pb)}, red solid lines 
   denote results of the TDHF calculations, blue dotted lines denote results 
   of the Langevin calculation \cite{Zagrebaev(2008)}, and green open 
   triangles connected with dotted lines denote results of the CWKB 
   calculation \cite{Corradi(58Ni+208Pb)}. 
   }
   \label{FIG:NTCS(58Ni+208Pb)_pure}
\end{figure}
%&&&&&&&&&&&&&&&&&&&&&&&&&&&&&&&&&&&&&&&&&&&&&&&&&&

We next consider an approach based on Langevin-type equations of 
motion which has been originally developed for and applied to fission
dynamics \cite{Abe(1996),Frobrich(1998)} and has been recently 
extended to apply to multinucleon transfer reactions 
\cite{Zagrebaev(2005),Zagrebaev(2007)1}.
We consider the \RE{58}{Ni}+{208}{Pb} reaction for which 
an application of the Langevin approach has been reported in 
Ref.~\cite{Zagrebaev(2008)}.

In the Langevin approach, multinucleon transfer processes are treated
as sequential processes of single-nucleon transfers. In the theory,
an empirical parameter describing nucleon transfer rate is introduced.
Figure~\ref{FIG:NTCS(58Ni+208Pb)_pure} shows cross sections 
for transfer channels of pure proton stripping without neutron 
transfer (left) and pure neutron pickup without proton transfer (right). 
Red filled circles denote measured cross sections, red solid lines denote 
results of our TDHF calculations, and blue dotted lines denote results 
of the Langevin approach reported in Ref.~\cite{Zagrebaev(2008)}. 

In the case of pure neutron pickup channels, ($0$p), the TDHF calculation
gives a better description than the Langevin theory for cross sections 
up to four-neutron transfer. The TDHF calculation overestimates the
cross sections for more than three neutrons, due to the quasi-fission
process at small impact parameter region as discussed in 
Sec.~\ref{Subsec(58Ni+208Pb)}. On the other hand, cross sections of 
pure proton stripping channels, ($0$n), are much better described by 
the Langevin theory than the TDHF, except for one-proton transfer channel.

%&&&&&&&&&&&&&&&&&&&&&&&&&&&&&&&&&&&&&&&&&&&&&&&&&&
\begin{figure}[t]
   \begin{center}
   \includegraphics[width=8.6cm]{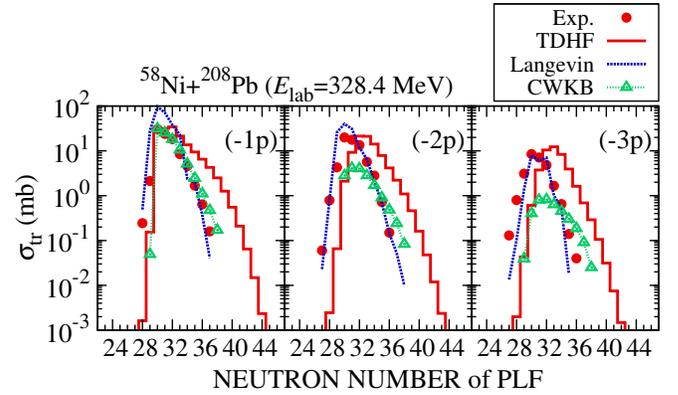}
   \end{center}\vspace{-3mm}
   \caption{(Color online)
   Cross sections for transfer channels of ($x$p) ($-3 \le x \le -1$) 
   for the \RE{58}{Ni}+{208}{Pb} reaction at $E_\mathrm{lab}=$ 
   328.4~MeV. The horizontal axis is the number of neutrons in the PLF. 
   Red filled circles denote measured cross sections 
   \cite{Corradi(58Ni+208Pb)}, red solid lines denote results of the 
   TDHF calculations, blue dotted lines denote results of the Langevin 
   calculation \cite{Zagrebaev(2008)}, and green open triangles 
   connected with dotted lines denote results of the CWKB calculation 
   \cite{Corradi(58Ni+208Pb)}.
   }
   \label{FIG:NTCS(58Ni+208Pb)_w_Zag}
\end{figure}
%&&&&&&&&&&&&&&&&&&&&&&&&&&&&&&&&&&&&&&&&&&&&&&&&&&

Figure~\ref{FIG:NTCS(58Ni+208Pb)_w_Zag} shows transfer cross 
sections for several proton stripping channels. Again, red filled circles 
denote measured cross sections, red solid lines denote results of our 
TDHF calculations, and blue dotted lines denote results of the Langevin
approach. For these channels, the Langevin calculation gives a much better
description for the transfer cross sections than the TDHF calculation.
We should, however, note that an adjustable parameter describing the 
nucleon transfer rate is introduced in the Langevin approach, while no
empirical parameter is introduced in the TDHF calculation once the 
Skyrme interaction is specified. In the calculation of the Langevin theory, 
evaporation effects are already included in the calculation.

For the \RE{58}{Ni}+{208}{Pb} reaction, an analysis using 
the CWKB theory has also been reported in 
Ref.~\cite{Corradi(58Ni+208Pb)}. In the CWKB theory, the 
multinucleon transfer processes are treated in a similar way to 
the GRAZING theory, evaluating statistically using single-nucleon 
transfer probabilities which are calculated by the first-order 
perturbation theory. In Figs.~\ref{FIG:NTCS(58Ni+208Pb)_pure} 
and \ref{FIG:NTCS(58Ni+208Pb)_w_Zag}, the CWKB cross 
sections are shown by green open triangles connected with dotted 
lines. In Ref.~\cite{Corradi(58Ni+208Pb)}, three results of cross 
sections have been reported: in a simple CWKB theory, adding 
the proton pair transfer effect, and taking account of evaporation 
effects in addition to the proton pair transfer effect. We show in 
these two figures the simplest version of the calculation without 
the proton pair transfer effect and the evaporation.

As seen from Fig.~\ref{FIG:NTCS(58Ni+208Pb)_pure}, the CWKB 
cross sections are very close to those of the TDHF except for 
$\Delta N \ge 5$ in the (0p) panel. 
In Fig.~\ref{FIG:NTCS(58Ni+208Pb)_w_Zag}, the CWKB cross 
sections are seen to be too small as the number of transferred 
protons increases. In Ref.~\cite{Corradi(58Ni+208Pb)}, proton 
pair transfer processes are introduced and added to the simple 
CWKB cross sections to examine the effect as a possible origin of 
the discrepancy.

\section{SUMMARY AND CONCLUSIONS}{\label{Sec:Summary_and_Conclusions}}

In the present paper, we have reported fully microscopic 
calculations for the multinucleon transfer reactions in the 
time-dependent Hartree-Fock (TDHF) theory.
We performed calculations for the reactions of 
\RE{40,\,48}{Ca}+{124}{Sn} at $E_\mathrm{lab}=$ 170, 174~MeV, 
\RE{40}{Ca}+{208}{Pb} at $E_\mathrm{lab}=$ 235, 249~MeV, and 
\RE{58}{Ni}+{208}{Pb} at $E_\mathrm{lab}=$ 328.4~MeV, 
for which multinucleon transfer cross sections have been 
measured experimentally \cite{Corradi(40Ca+124Sn),
Corradi(48Ca+124Sn),Szilner(40Ca+208Pb)2,Corradi(58Ni+208Pb)}.
We use a projection operator method \cite{Projection} to calculate 
the transfer probabilities as functions of impact parameter from the 
wave function after collision. From the reaction probabilities, 
we evaluate cross sections for various transfer channels.

The systems we have calculated, \RE{40,\,48}{Ca}+{124}{Sn}, 
\RE{40}{Ca}+{208}{Pb}, and \RE{58}{Ni}+{208}{Pb} show different
behaviors in the multinucleon transfer processes characterized 
by the $N/Z$ ratios of the projectile and the target, and by
the product of the charge numbers, $Z_\mathrm{P}Z_\mathrm{T}$.

In the collisions with different $N/Z$ ratios between the
projectile and the target (\RE{40}{Ca}+{124}{Sn}, 
\RE{40}{Ca}+{208}{Pb}, and \RE{58}{Ni}+{208}{Pb}), 
we find a fast transfer of a few nucleons when the impact
parameter is sufficiently large. The nucleons are transferred
towards the direction of charge equilibrium expected from the
$N/Z$ ratios of the projectile and the target. This means that 
protons and neutrons are transferred in the opposite directions.
When the $N/Z$ ratios are almost equal between the projectile
and the target (\RE{48}{Ca}+{124}{Sn}), we find a few
nucleons are exchanged symmetrically.

As the impact parameter decreases, a neck is formed at the
contact of two nuclei. Then the transfer process proceeds
in two steps. At the beginning of the reaction before the 
formation of the neck, a few nucleons are transferred
in the same way as described above. After forming the neck,
the transfer of a number of nucleons occurs as a result of 
the neck breaking when two nuclei dissociate. Since both 
protons and neutrons in the neck are transferred simultaneously
in this mechanism, protons and neutrons are transferred 
in the same direction.

As the charge number product $Z_\mathrm{P}Z_\mathrm{T}$ 
increases, there appears an impact parameter region where a
thick neck is formed. We consider reactions in this impact
parameter region correspond to the quasi-fission. As mentioned 
above, both protons and neutrons are transferred in the same 
direction when the neck is broken at the time of dissociation.
A large energy transfer is also accompanied from the 
nucleus-nucleus relative motion to the internal excitations.

Comparisons with measured cross sections show that the TDHF 
calculations describe cross sections reasonably well for transfer 
processes of a few nucleons between the projectile and the
target. As the number of exchanged nucleons increases, the 
agreement becomes less accurate. When more than a few 
protons are transferred, cross sections as functions of the 
transferred neutron number show a peak at the neutron 
number more than that in the measurements. The magnitude 
of the calculated cross sections becomes too small compared with 
the measurements. This discrepancy is expected to be, to some extent, 
resolved when we introduce nucleon evaporation effects in our 
calculations.

We have compared transfer cross sections of the TDHF 
calculations with those by other theories. We find that results of 
the TDHF calculations are rather close to those of direct reaction 
model calculations such as GRAZING and Complex WKB. We 
should note that the Skyrme Hamiltonian used in the TDHF 
calculation is entirely determined from the ground state 
calculations and there is no parameter introduced to describe 
nuclear dynamics. We thus conclude that the fully microscopic 
TDHF theory may describe the multinucleon transfer cross sections 
in the quality comparable to existing direct reaction theories.

To increase the reliability of our calculation, an inclusion of 
evaporation effects is apparently important. Since the evaporation
processes take place in much longer time scale than the reaction
mechanism shown here, it is not realistic to achieve the calculation
within the TDHF theory. Instead, we are considering to estimate them 
using a statistical model, putting excitation energy of the final 
fragments calculated in the TDHF theory as inputs.

An inclusion of correlation effects beyond the mean-field
approximation is also of much significance. Attempts beyond the 
mean-field have been seriously undertaken recently in realistic 
calculations. For example, in Ref.~\cite{Simenel(BV2011)}, 
calculations based on the Barian-V{\'e}n{\'e}roni prescription 
have been reported. In Refs.~\cite{Ayik(2008),Ayik(2009),
Washiyama(2009)SMF,Yilmaz(2011)}, a stochastic effects are 
included in the mean-field dynamics.

One of the most important correlations in nuclear low-energy 
dynamics is the pairing correlation. A full solution of the time-dependent
Hartree-Fock-Bogoliubov (TDHFB) theory \cite{Hashimoto(2007),
Avez(2008),Stetcu(2011),Hashimoto(2012)} would provide a 
satisfactory description for it. At present, however, there have not 
yet been reported realistic TDHFB calculations of heavy-ion 
collisions. A simplified method to include the pairing effects 
in the TDHF dynamics has also been discussed 
\cite{BlockiFkocard(1976),Ebata(2010),Scamps(2012)1,
Ebata(2012),Scamps(2012)2}. Extending the present analysis 
including the pairing correlation will certainly bring more satisfactory 
description and provide reliable understanding for the multinucleon 
transfer reactions.

\begin{acknowledgments}
We would like to thank Professor L.~Corradi for giving us the experimental 
data. K.S. greatly appreciates S.~Ebata for stimulating discussions 
and encouragements. Discussions during the YIPQS2011 symposium on 
``Frontier Issues in Physics of Exotic Nuclei 2011" and the international 
workshop on ``Dynamics and Correlations in Exotic Nuclei 2011" 
(DCEN2011), were useful to complete this work. Numerical calculations 
for the present work have been carried out on the T2K-Tsukuba 
supercomputer at the University of Tsukuba and on the HITACHI 
SR16000 Model M1 supercomputer at the Hokkaido University. 
This work is supported by the Grants-in-Aid for Scientific Research 
Nos. 21340073, 23340113, and 23104503.
\end{acknowledgments}

\appendix

\section{Average and fluctuation of nucleon number}{\label{App:average}}

We summarize formula for the average and the fluctuation of the 
nucleon number belonging to the PLF or to the TLF. Using the 
probability distribution $P_n$, defined by Eq.~(\ref{Definition_of_P_n}), 
the average and the fluctuation of the nucleon number belonging 
to the PLF are given by
\begin{eqnarray}
\nonum\\[-6mm]
& \displaystyle{ \bigl<\hat{N}_{V_\mathrm{P}} \bigr> = 
\sum_{n=0}^N n \,P_n, }& \\
& \displaystyle{ \sigma = 
\sqrt{\sum_{n=0}^N \Bigl( 
n-\bigl< \hat{N}_{V_\mathrm{P}} \bigr> 
\Bigr)^{\hspace{-0.2mm}2} \hspace{-0.5mm}P_n} }.&
\end{eqnarray}

The average and the fluctuation of the nucleon number belonging
to the PLF may also be calculated directly from the orbitals without
projection procedure. The average number is calculated from the density, 
$\rho(\bvecr) = \sum_{i=1}^N\sum_\sigma 
\psi_i^*(\bvecr,\sigma)\psi_i(\bvecr,\sigma)$,
integrating over the spatial region $V_\mathrm{P}$:
\begin{equation}
\bigl<\hat{N}_{V_\mathrm{P}} \bigr> = 
\int_{V_\mathrm{P}}\hspace{-2mm} d\bvecr \; \rho(\bvecr).
\end{equation}
The fluctuation can also be directly calculated from the single-particle
orbitals by
\begin{eqnarray}
\sigma & \equiv & \sqrt{\bigl< \bigl( \hat{N}_{V_\mathrm{P}} \bigr)^2 \bigr> 
- \bigl< \hat{N}_{V_\mathrm{P}} \bigr>^2} \nonum\\[1mm]
&=& \sqrt{ \bigl<\hat{N}_{V_\mathrm{P}} \bigr> 
- \sum_{i,j=1}^N \abs{\bracket<\psi_i|\psi_j>_{V_\mathrm{P}}}^2 }.
\label{fluctuation_sigma}
\end{eqnarray}

\end{document}